\documentclass{aa}  
\usepackage{natbib}
\usepackage{graphicx}
\usepackage{txfonts}
\usepackage{hyperref}
\usepackage{amsmath}
\usepackage{nccmath}
\usepackage{subfig}
\usepackage{fixltx2e}
\usepackage[utf8]{inputenc}
\usepackage{color}
\usepackage{ulem}
\begin{document} 
  \title{KMOS LENsing Survey (KLENS) : morpho-kinematic analysis of star-forming galaxies at $z \sim 2$\thanks{Based on KMOS observations made with the European Southern Observatory VLT/Antu telescope, Paranal, Chile, collected under the programme ID No.~095.A-0962(A)+(B).} 
  }

  \author{M. Girard\inst{1}
  \and 
  M. Dessauges-Zavadsky\inst{1}
  \and 
  D. Schaerer\inst{1,2}
     \and
 M. Cirasuolo\inst{3,4}  %\fnmsep
      \and
  O. J. Turner\inst{4,3}  %\fnmsep   
        \and
        A. Cava\inst{1}
        \and
    L.~Rodr\'{i}guez-Mu\~{n}oz\inst{5}  %\fnmsep   
      \and
  J. Richard\inst{6}
        \and
        P. G. P\'{e}rez-Gonz\'{a}lez\inst{7}
          }

  \institute{Observatoire de Gen\`{e}ve, Universit\'{e} de Gen\`{e}ve, Ch. des Maillettes 51, Versoix 1290, Switzerland\\
              \email{marianne.girard@unige.ch}
         \and
            CNRS, IRAP, 14 Avenue E. Belin, 31400 Toulouse, France
            \and
            European Southern Observatory, Karl-Schwarzschild-Str. 2, 85748 Garching b. M\"{u}nchen, Germany
           \and 
           SUPA, Institute for Astronomy, University of Edinburgh, Royal Observatory, Edinburgh EH9 3HJ
             \and
                Dipartimento di Fisica e Astronomia, Universit\`{a} di Padova, vicolo dellOsservatorio 2, 35122 Padova, Italy
           \and
          CRAL, Observatoire de Lyon, Universit\'{e} Lyon 1, 9 Avenue Ch. Andr\'{e}, 69561 Saint Genis Laval Cedex, France
            \and    
          Departamento de Astrof\'{i}sica y CC. de la Atm\'{o}sfera, Universidad Complutense de Madrid, E-28040 Madrid, Spain
             }

  \date{Received [...], 2017; accepted [...] }

%%%%%%%%%%%%%%%%%%%%%%%%%%%%%%%%%%%%%%%%
%           ABSTRACT
%%%%%%%%%%%%%%%%%%%%%%%%%%%%%%%%%%%%%%%%

  \abstract{
  We present results from the KMOS lensing survey (KLENS), which is exploiting gravitational lensing to study the kinematics of 24 star forming galaxies at $1.4<z<3.5 $ with a median mass of $\rm log(M_\star/M_\odot)=9.6 $ and median star formation rate (SFR) of $\rm 7.5 \, M_\odot \, yr^{-1} $. We find that 25\% of these low-mass/low-SFR galaxies are rotation dominated, while the majority of our sample shows no velocity gradient. When combining our data with other surveys, we find that the fraction of rotation dominated galaxies increases with the stellar mass, and decreases for galaxies with a positive offset from the main sequence (higher specific star formation rate). We also investigate the evolution of the intrinsic velocity dispersion, $\sigma_0$, as a function of the redshift, $z$, and stellar mass, $\rm M_\star$, assuming  galaxies in quasi-equilibrium (Toomre Q parameter equal to 1). From the $z-\sigma_0$ relation, we find that the redshift evolution of the velocity dispersion is mostly expected for massive galaxies ($\rm log(M_\star/M_\odot)>10 $). We derive a $\rm M_\star-\sigma_0$ relation, using the Tully-Fisher relation, which highlights that a different evolution of the velocity dispersion is expected depending on the stellar mass, with lower velocity dispersions for lower masses, and an increase for higher masses, stronger at higher redshift. The observed velocity dispersions from this work and from comparison samples spanning $0<z<3.5$ appear to follow this relation, except at higher redshift ($z>2$), where we observe higher velocity dispersions for low masses ($\rm log(M_\star/M_\odot)\sim 9.6$) and lower velocity dispersions for high masses ($\rm log(M_\star/M_\odot)\sim 10.9$) than expected. This discrepancy could, for instance, suggest that galaxies at high redshift do not satisfy the stability criterion, or that the adopted parametrisation of the specific star formation rate and molecular properties fail at high redshift.
  }

  \keywords{galaxies: high redshift -- galaxies: evolution -- galaxies:kinematics and dynamics
            }

\maketitle

%%%%%%%%%%%%%%%%%%%%%%%%%%%%%%%%%%%%%%%%
%           INTRODUCTION
%%%%%%%%%%%%%%%%%%%%%%%%%%%%%%%%%%%%%%%%

\section{Introduction}

The kinematics of star-forming galaxies (SFGs) are known to provide key information about their fundamental properties. Studies of SFGs in the local Universe have shown that rotation curves are a powerful tool to understand their structure, mass distribution, dynamics, formation, and interactions 
\citep[e.g.][]{Persic1996,Sofue2001,Glazebrook2013}. Several scaling relations between these physical properties have now been established, for example the empirical relation between luminosity or mass and rotation velocity of the disc, known as the Tully-Fisher relation
\citep[e.g.][]{Tully1977,Pizagno2007,Courteau2007,Reyes2011}. By studying the evolution of such scaling relations we can gain insight into the physical conditions of high redshift galaxies.

Using near-infrared instruments, it has become possible to study the kinematics of galaxies at a particularly interesting epoch, around $z \sim 2$, when the cosmic star formation rate (SFR) density is at its peak and the mass assembly of galaxies is rapid 
\citep[e.g.][]{Madau1998,Perez-Gonzalez2005,Li2008}. Several large surveys (e.g.  HiZELS, SINS/zC-SINF, MASSIV, KMOS$^{\rm 3D}$, KROSS) of spatially-resolved ionized gas kinematics (including velocity gradients, velocity dispersion profiles, dynamical masses), star formation, and physical properties (metallicity gradients, excitation) of SFGs at $0.8 < z < 2.7$ have been undertaken with the integral field unit (IFU) spectrographs SINFONI/VLT, KMOS/VLT, and OSIRIS/Keck II 
\citep[e.g.][]{ForsterSchreiber2009,Epinat2012,Sobral2013,Wisnioski2015,Stott2016}.

Overall, these surveys reveal that a majority (60-70\%) of the resolved high redshift galaxies are rotation dominated, while a minority of galaxies consist of merging systems and more compact, velocity dispersion dominated objects. The morphologies of these objects are observed to be more compact, clumpy or irregular than at low redshift even if a rotating disc is observed.
An increase of the intrinsic velocity dispersion with redshift is observed and is believed to be related to the higher gas fractions observed at high redshift 
\citep[e.g.][]{Law2009,Daddi2010,Geach2011,Tacconi2013,Saintonge2013,Sargent2014,Genzel2015,Dessauges-Zavadsky2015,Dessauges-Zavadsky2017}. In addition to larger gas reservoirs, this increase of random motions with redshift could be due to different effects such as disc instabilities  \citep[e.g.][]{Bournaud2016,Stott2016} or an increase of the accretion efficiency \citep{Law2009}. 
The Tully-Fisher relation at high redshift is still debated, since \citet{Harrison2017} find a relation at $z \sim 0.9$ consistent with the local universe, while a redshift evolution from $z=0$ is obtained for two samples at $z \sim 0.9$ and $z\sim2.5$ by \citet{Ubler2017}.

Clearly, these near-infrared studies have revolutionized our understanding of SFGs at $z\sim 2$. However, the picture is still incomplete, as the spatially-resolved surveys at high redshift remain limited to fairly massive galaxies ($\rm log (M_\star/M_\odot) \gtrsim 10$) with large star formation rates ($\rm SFR \gtrsim 30 \, M \, yr^{-1}$), whereas the majority of galaxies at this epoch have a lower mass and SFR. Indeed, the characteristic mass $\rm M^\star$ is known to be $\rm log(M^\star/M_\odot) \sim 10.3$ at $z \sim 2$ \citep{Ilbert2013} and both the H$\alpha$ and infrared luminosity functions show that the characteristic $\rm SFR^\star$ is between 30-$\rm60 \, M_\odot \, yr^{-1}$ \citep[][]{Sobral2013b,Burgarella2013}. Studying galaxies with $\rm log(M_{\star}/M_{\odot}) < 10 $ at $ z\sim$ 1.5-2, the progenitors of the Milky Way, is essential to gain further evolutionary insights
\citep[][]{Behroozi2013,Haywood2013}.

Recent studies have investigated low mass galaxies at $ z<2$ with VLT/MUSE and VLT/KMOS \citep{Contini2016, Swinbank2017}. \citet{Gnerucci2011} and \citet{Turner2017} have used deep observations with VLT/SINFONI and VLT/KMOS to obtain a sample of low mass galaxies at $ z\sim3.5$. Another way to extend these studies to more numerous low mass and/or low SFR galaxies is by exploiting gravitational lensing, which allows us to probe below the knee of the luminosity function with current instrumentation.
Already several studies have used this powerful tool to study kinematics
\citep{Jones2010, Livermore2015, Leethochawalit2016, Mason2017}.

These lensed and deep surveys of low mass SFGs at high redshift confirm a high intrinsic velocity dispersion ($\rm \sigma \gtrsim 50 \, km \, s^{-1}$) until at least a redshift of 3.5 \citep{Gnerucci2011, Livermore2015, Leethochawalit2016, Mason2017, Turner2017}. Simulations have also shown that an increase of the dispersion is expected for Milky Way progenitors at least until a redshift of 1.2 \citep{Kassin2014}.

Overall, these low mass SFG surveys report a rotation dominated fraction significantly lower ($< 50\%$) than previous studies based on more massive galaxies. Recent simulations of galaxies in the local Universe have shown that the fraction of rotation dominated galaxies is expected to increase with stellar mass \citep{El-Badry2017}. In this study, at $\rm log(M_\star/M_\odot)~<~8$, very few galaxies are supported by rotation and none of them have a rotation curve which is flattened by rotation. At $\rm 8~<~log(M_\star/M_\odot)~<~10$, both dispersion and rotation supported galaxies are obtained, and at $\rm 10<log(M_\star/M_\odot) < 11$, all the galaxies form a disc dominated by rotation. Observations in the local Universe are in good agreement with this picture, since a large diversity in the morphology and kinematics of low mass galaxies has been reported 
\citep[e.g.][]{Walter2008,Ott2012,Roychowdhury2013,Simons2015}.

Nevertheless, the mass dependence on the rotation dominated fraction and intrinsic velocity dispersion needs to be established from observations at high redshift, prompting us to carry out the KMOS LENsing Survey (KLENS). The goal of this survey is to extend existing near-infrared spectroscopic surveys using gravitationnal lensing to more numerous and typical galaxies  at $z\sim 2$ with low mass ($\rm log(M_\star/M_\odot)\sim 9.5 $) and/or SFR ($\rm SFR <30 \, M_\odot \, yr^{-1}$).

The paper is organized as follows. Section \ref{section2} describes our sample selection, observations, data reduction and comparison samples. In Sect. \ref{section3}, we explain   measurements of the integrated properties and the SED fitting technique used to derive the stellar mass and the star formation rate. The morphological analysis of  our sample is described in Sect. \ref{section4}. Section \ref{section5} presents the kinematics and classification adopted in our work. Section \ref{section6} presents a discussion on the fraction of rotation dominated galaxies and the evolution of the kinematic properties, specifically the intrinsic velocity dispersion as a function of the redshift and stellar mass. We finally present our conclusions in Sect. \ref{Conclusions}.

In this paper, we use a cosmology with H$_0=70$~km~s$^{-1}$~Mpc$^{-1}$, $\Omega_M=0.3$, and $\Omega_{\Lambda}=0.7$. When using values calculated with the initial mass function (IMF) of \citet{Salpeter1955}, we correct by a factor of 1.7 to convert to a \citet{Chabrier2003} IMF.

%%%%%%%%%%%%%%%%%%%%%%%%%%%%%%%%%%%%%%%%
%           Section2
%%%%%%%%%%%%%%%%%%%%%%%%%%%%%%%%%%%%%%%%%

\section{Observations and data reduction}
\label{section2}

\subsection{Sample selection and observations}
\label{section2.1}

We have selected our targets from two galaxy clusters from  the CLASH programme (\citeauthor{Postman2012} \citeyear{Postman2012}) and included in the Herschel Lensing Survey (HLS; \citeauthor{Egami2010} \citeyear{Egami2010}) : MACS1206-08 and AS1063. For each cluster, the existing imaging, accurate magnification maps, and near-ultraviolet to far-infrared SED fits for all objects are available in the \mbox{RAINBOW} database \citep{PerezGonzalez2008, Barro2011a, Barro2011b}, allowing us to select in an automated fashion galaxies based on their magnification, redshift, stellar mass, star formation rate, color-selection, and infrared luminosity.
The criteria for our galaxy selection were~:
\begin{itemize}
\item A photometric redshift between 1.3 and 3.5 to detect the H$\alpha$ or [OIII] emission lines in the H or K bands (Fig. \ref{hist_klens}, left panel);
\item An observed SFR higher than $ 10 \, \mathrm{M}_\odot \, \mathrm{yr}^{-1}$ to ensure a 5$\sigma$ detection of H$\alpha$ or [OIII];
\item A SFR corrected for the magnification (SFR/$\mu$) below $30~\mathrm{M}_\odot \, \mathrm{yr}^{-1}$ or a stellar mass  corrected for the magnification ($\rm M_{\star}/\mu$) below $\rm 10^{10} \, M_{\odot}$, to extend existing near-infrared spectroscopic surveys below the current observational limits;
\item A magnification of $\mu \sim$ 1.5-4 and therefore galaxies located away from the critical lines to avoid effects caused by differential magnification within the object. In these seeing-limited data, the shear introduced by weak-lensing is also negligible and does not affect the kinematics. 
\end{itemize}

We have initiated KLENS in 2015 in P95 with the K-band Multi Object Spectrograph 
\citep[KMOS;][]{Sharples2013}. KMOS is an instrument with 24 arms of $14 \times 14$ spaxels. Each spaxel has $0.2" \times 0.2"$ which gives a global field of view of $2.8" \times 2.8"$ for each arm. 
Observations were carried out in the H and K bands, which have a typical spectral resolution of $R \sim 4000$ and $R \sim 4200$, respectively. Each pointing had an exposure time of 300s and we used an object-object-sky-object-object dither pattern. The sky frames were obtained by applying an offset to a clear sky position. The observations were taken in good conditions with a seeing around 0.6" in H and K bands. The total on-source exposure time in the H band is 2.3h for both clusters. In the K band, the targets have been observed during 8h and 10h on-source for MACS1206-08 and AS1063, respectively.

Sixty different galaxies in MACS1206-08 and AS1063 have been observed. For these 60 galaxies, 19 at $z \sim 2.2$  were observed in both H and K bands, 21 at $z \sim 3.0$ were observed only in the K band and 20 at $z \sim 1.5$ only in the H band. Figure \ref{hist_klens} (left panel) presents the photometric redshifts of the observed galaxies.

\subsection{Data reduction}

The data reduction was performed primarily using the official KMOS ESOREX/SPARK pipeline \citep{Davies2013}, with the addition of custom Python scripts as described in detail in \citet{Turner2017}. Each object frame (i.e. containing the science objects) was reduced individually using the sky frame nearest to it in time. The subtraction of the sky background was enhanced by using the SKYTWEAK optional routine within the ESOREX/SPARK pipeline. This routine performs a flux scaling of the individual OH sky lines families to match the data, as well as a spectral shift to account for any wavelength miscalibration between object and sky frame \citep{Davies2013}. 

Each sky-subtracted object frame was flux calibrated by using observations of a standard star taken the same night (before or after the science observations). The individual 300sec exposures where then stacked using a clipped average, providing a flux and wavelength calibrated data cube for every object. These cubes were used to create velocity maps and to extract one-dimensional spectra for the analysis presented here.

\subsection{Comparison samples}
\label{comparisonsamples}

In what follows, we summarize the different surveys we use for comparison in this work. We extract, from all the surveys mentioned below, the stellar mass, SFR, both corrected for \citet{Chabrier2003} IMF, the redshift, velocity dispersion and kinematic classification, when publicly available. 

\citet{Livermore2015} present the kinematics of 12~gravitationally lensed galaxies at $ 1 < z < 4 $ with $8.6~<~\mathrm{log}(\mathrm{M}_{\star}/\mathrm{M}_\odot)~<~10.8 $ (median of 9.4) and SFRs between 0.8 and 40 M$_{\odot}$yr$^{-1}$. \citet{Leethochawalit2016} discuss the kinematics of 11 lensed galaxies at $ 1.4 < z < 2.5 $ with a mass range of $9.0 <\mathrm{log}(\mathrm{M}_{\star}/\mathrm{M}_\odot) < 9.6 $ and SFRs between 5 and 250 M$_{\odot}$yr$^{-1}$. The KMOS Lens-Amplified Spectroscopic Survey (KLASS) by \citet{Mason2017} uses gravitationnal lensing to study the kinematics of 25 galaxies at $ 0.7 < z < 2.3 $ with  $7.8< \mathrm{log}(\mathrm{M}_{\star}/\mathrm{M}_\odot)~<~10.5 $ (median of 9.5) and SFRs between 0.1 and 110 M$_{\odot}$yr$^{-1}$. We use these three lensed surveys mainly to increase the number of galaxies in our sample, since these lensed galaxies have similar redshifts, stellar masses, and star formation rates to galaxies in KLENS. These three surveys, in addition to KLENS, will be referred to as the lensed surveys.

Moreover, for comparison at lower and higher redshifts we use the data of \citet{Contini2016} obtained with MUSE (MUSE Hubble Deep Field South - HDFS) and \citet{Turner2017} obtained with KMOS (KMOS Deep Survey - KDS). \citet{Contini2016} study 28 galaxies at $0.2 < z < 1.4 $ in the mass range of $7.9< \mathrm{log}(\mathrm{M}_{\star}/\mathrm{M}_\odot) < 10.8 $ (median of 9.1) and SFRs between 0.01 and 80 M$_{\odot}$yr$^{-1}$. \citet{Turner2017} present the kinematics of 38 resolved field  galaxies  at  $ 3.0 < z < 3.8 $ in the stellar mass range $9.0< \mathrm{log}(\mathrm{M}_{\star}/\mathrm{M}_\odot) < 10.5 $ (median of 9.7) and SFRs between 4 and 175~M$_{\odot}$yr$^{-1}$. 

We finally also consider the commonly used surveys MUSE-KMOS \citep{Swinbank2017}, AMAZE \citep{Gnerucci2011}, MASSIV \citep{Epinat2012},  KMOS$^{\rm 3D}$ \citep{Wisnioski2015}, and SINS \citep{ForsterSchreiber2009}. All these surveys have been summarized in the recent work of \citet{Turner2017}.

%__________________________________________________________________

%Figure klens histogram
   
      \begin{figure*}
   \centering
   \subfloat{\includegraphics[scale=0.275]{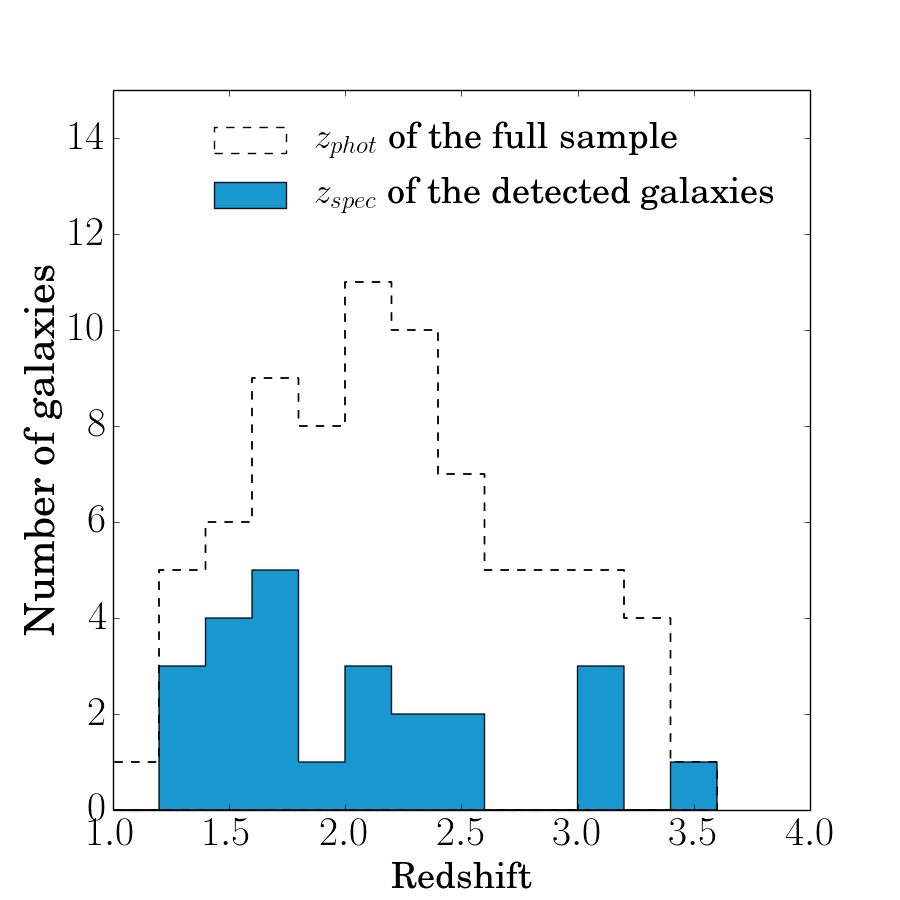}}
  \subfloat{\includegraphics[scale=0.275]{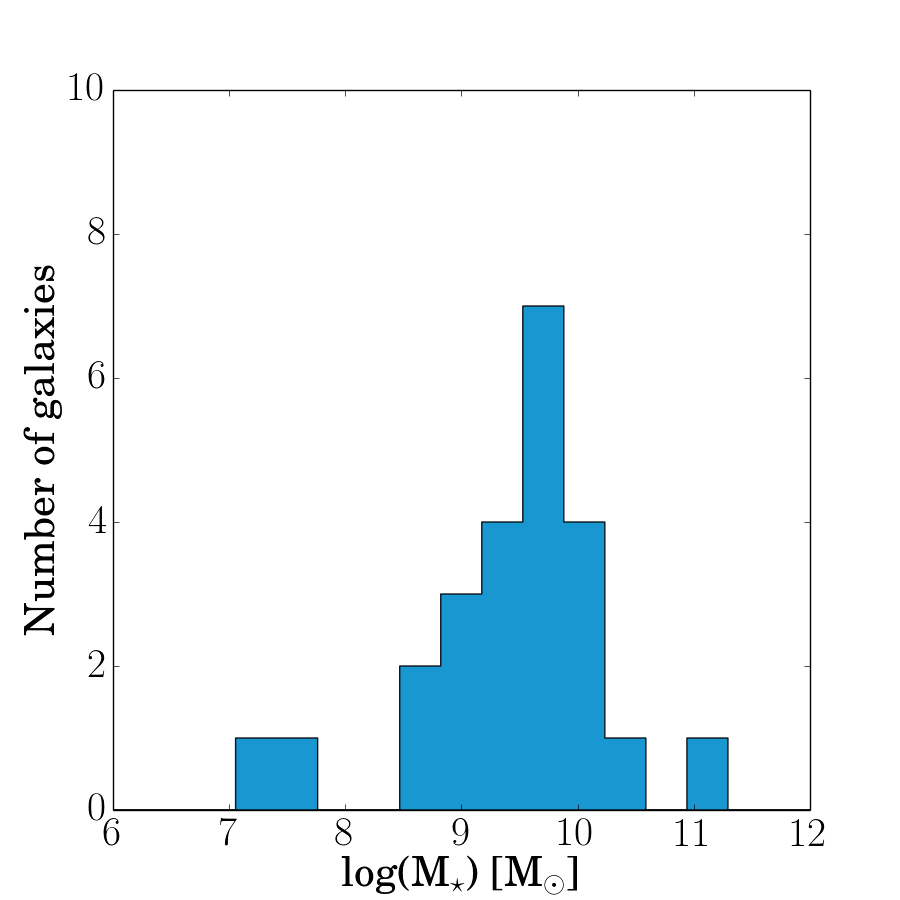}}
    \subfloat{\includegraphics[scale=0.275]{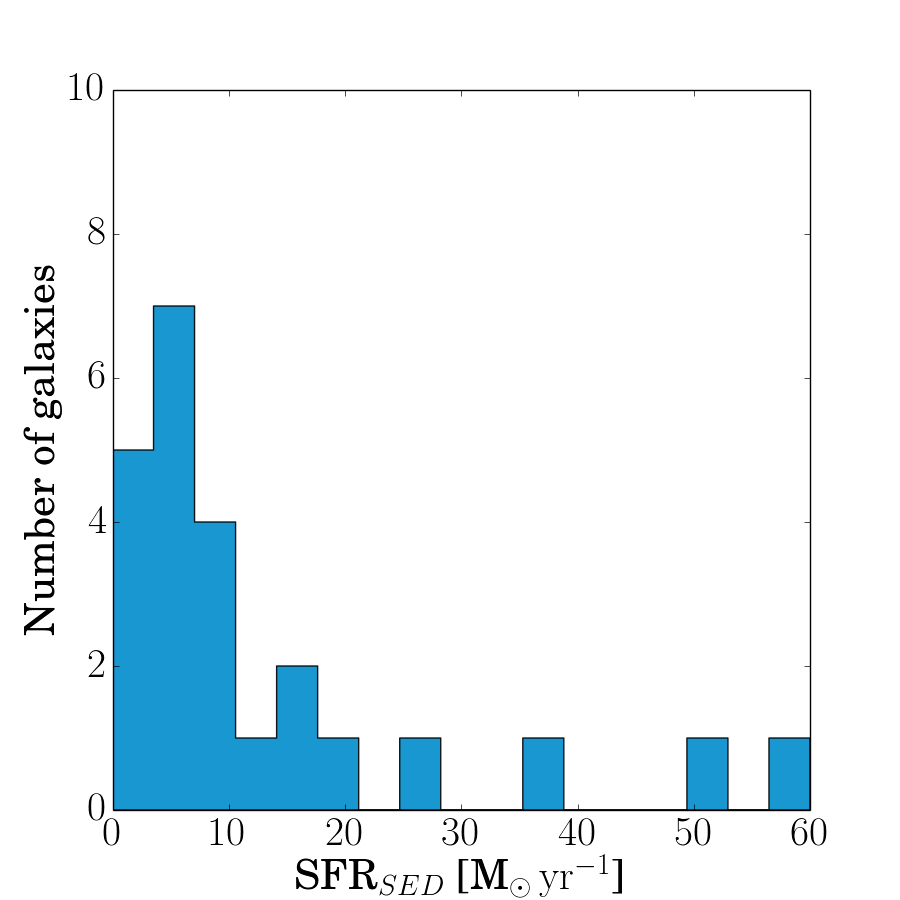}}

      \caption{Histograms of the redshift (left panel), stellar mass (left panel), and star formation rate derived from the SED fitting (right panel) for the galaxies detected in our sample. The values are all lensing-corrected.}
         \label{hist_klens}
   \end{figure*}

%__________________________________________________________________

%%%%%%%%%%%%%%%%%%%%%%%%%%%%%%%%%%%%%%%%
%           Section3
%%%%%%%%%%%%%%%%%%%%%%%%%%%%%%%%%%%%%%%%

\section{Measurements}
\label{section3}
\subsection{Detection of emission lines}

We have set up a systematic way to detect emission lines in datacubes since the spectroscopic redshift of most of our targets was unknown. We initially use the Line Source Detection and Cataloguing software 
\citep[LSDcat;][]{Herenz2017}. This software was first developed for large field of view IFUs like VLT/MUSE, but is also applicable to datacubes from VLT/KMOS. LSDcat allows us to detect the objects in our sample with the highest signal-to-noise. However, the noisy datacubes we obtained prevent us from detecting weak emission lines with this technique. 

Most of the detections have been made by fixing a central spaxel, corresponding spatially to the peak of HST emission, and creating a unique spectrum by summing all the spaxels inside a radius of 2 (0.4")  and 5 (1.0") spaxels. The skylines are also masked after the creation of these spectra. In this way, we get two spectra for each target to inspect visually. The 0.4" spectrum allows us to detect emission lines from more compact galaxies and the 1.0" spectrum allows us to detect more diffuse or extended galaxies. 

We detect H$\alpha$ or [OIII] emission lines in 24 targets out of 60 (i.e. 40\% detection rate) from the galaxy clusters AS1063 and MACS1206-08. Two of these detected targets are observed in both H and K bands, 17 are detected in the H band, and 5 are detected in the K band only. We detect in addition to H$\alpha$ and/or [OIII] the [NII] emission line in one galaxy and H$\beta$ in three galaxies, but these lines are faint and only visible in the integrated spectra. The detected emission lines are summarized in Table~\ref{table_integrated_properties}.

We do not detect any emission line for the remaining 36 targets. The median H-band magnitude of targets with non-detections is 23.4 compared to 23.0 for those with detections, which means that our low detection rate is not due to the detection limit. The low detection rate is likely due to a combination of various other effects. First, the strong and numerous skylines present in H and K bands which can be at the same position as the emission lines. Second, we do not have any detection longwards $2.3 \, \mathrm{\mu m} $ in the K band, even for the galaxies which have a known spectroscopic redshift and where we expect a detection. This part of the K band shows a high level of noise in all our datacubes and it is probably due to the lower transmission towards the edges of the filters and the thermal noise of the instrument.
As a consequence, the fractions of H and K bands lost are about $\sim 20\% $ and $\sim 60\%$ respectively, which cannot explain completely our detection rate of 40\%. 
It may also be that the photometric redshifts are not always accurate and therefore the emission lines are possibly outside the H and K bands. Finally, the H$\alpha$ or [OIII] lines of the sources are perhaps weaker than expected, and hence undetected.

\subsection{Redshift}

We have determined spectroscopic redshifts using the wavelength of [OIII] or H$\alpha$ emission lines from the one-dimensional spectra (Table \ref{table_integrated_properties}). We obtain a median difference of 0.18 between $z_{spec}$ and $z_{phot}$. When only one emission line is detected for a galaxy, we take the most likely redshift according to the probability density function (PDF). When we are not able to determine the redshift from the PDF, we perfom a SED fitting with both redshifts corresponding to [OIII] and H$\alpha$ emission lines and take the redshift for which we obtain the best fit. Figure \ref{hist_klens} (left panel) shows the redshift distribution of our sample.

The sources AS1063-1321 and AS1063-1152 are known to be multiple images of the same galaxy \citep{Karman2015}. We measure a redshift of $z=1.428$ which is in good agreement with  \citeauthor{Karman2015} (\citeyear{Karman2015}; source 46, $z=1.428$),
\citeauthor{Richard2014} (\citeyear{Richard2014}; source 2, $z=1.429$),
\citeauthor{Balestra2013} (\citeyear{Balestra2013}; source 6a, $z=1.429$),
and \citeauthor{Johnson2014} (\citeyear{Johnson2014}; source 6, $z=1.429$).

The redshift $z=2.580$ measured for the source AS1063-942 is also consistent with the redshift determined by \citeauthor{Karman2015} (\citeyear{Karman2015}; source 48, $z=2.577$).

The source AS1063-885 has been studied in detail by \citet{Vanzella2016} and \citet{Vanzella2017}. They report in their work that this object is extremely blue and young, showing characteristics similar to a proto-globular cluster. They find a redshift of $z=3.1169$, which is consistent with the redshift of $z=3.1171$ we measure in this work.

The sources MACS1206-1472 and MACS1206-820 are also known as multiple images of the same galaxy \citep{Zitrin2012}. The redshift of 3.03 reported is also in good agreement with our measurement of $z=3.038$.

%__________________________________________________________________

%TABLE Integrated properties
\begin{sidewaystable*}
\caption{Integrated properties}             
\label{table_integrated_properties}      
\centering          
\begin{tabular}{l c c c c c c c c c }     
\hline\hline       
Objects & Coordinates J2000 &   $z_{phot}$ & $z_{spec}$ & Lines & $\mu$\tablefootmark{a} &  log(M$_\star$)\tablefootmark{b}  & SFR$_{H\alpha}$\tablefootmark{b} & SFR$_{SED}$\tablefootmark{b} & $\Delta$SFR\tablefootmark{c}
\\
& RA - Dec & & & & & [M$_{\odot}$] & [M$_{\odot} \, \mathrm{yr}^{-1}$] &  [M$_{\odot} \, \mathrm{yr}^{-1}$] 
\\ 
\hline

AS1063-1816 & $342.1863831 \, -44.5211345$ & 1.52 & 1.2277  &  H$\alpha$  & $4.52 \pm 0.12 $& $ 9.24_{-0.09}^{+0.07}$ & $0.90 \pm 0.07$ & $1.2_{-0.1}^{0.2} $& $ -0.40$\\
AS1063-658 &    $342.1955084 \, -44.5470291$ & 1.48 &1.3528   & H$\alpha$ & $1.51  \pm 0.03  $& $9.95_{-0.09}^{+0.04}$ & $2.16 \pm 0.09 $& $3.9_{-0.5}^{+0.3} $& $-0.59$ \\
AS1063-1611 & $342.1790680 \, -44.5235948$ & 1.70 &  1.3971  &  H$\alpha$  & $3.84 \pm 0.11 $ & $8.91_{-0.02}^{+0.02}$ & $1.38\pm 0.09 $   & $15.6_{-0.8}^{+0.7}$ & 1.03\\
AS1063-1670 & $342.1991817 \, -44.5133024$ & 1.48  &1.4267  &  H$\alpha$  & $2.15 \pm 0.04$ & $11.04_{-0.01}^{+0.01}$ & $ 1.11 \pm 0.27 $& $4.7_{-0.1}^{+0.1}$  & $-1.08$\\
AS1063-1152\tablefootmark{d} & $342.1757472 \, -44.5324889$ &  1.61 & 1.4283  &  H$\alpha$  & $5.51 \pm 0.27$ & $9.83_{-0.04}^{0.10}$  & $0.49\pm 0.07  $& $2.1_{-0.5}^{+0.4} $& $-0.78$\\
AS1063-1321\tablefootmark{d} & $342.1741608 \, -44.5283941$ &  1.52 & 1.4285  &  H$\alpha$, [NII]  & $6.13 \pm 0.24$ & $9.57_{-0.03}^{+0.02}$ &$0.66\pm 0.11 $& $3.5_{-0.2}^{0.1}$ & $-0.31$\\
AS1063-1331 & $342.2084851 \, -44.5292163$ & 1.74 & 1.6372  &  H$\alpha$  & $2.14 \pm 0.04$ & $9.30_{-0.05}^{+0.03}$ & $1.24\pm0.11 $& $5.6_{-3.7}^{+1.6} $ & $-0.12$\\
AS1063-732 &    $342.2051886 \, -44.5444467$ & 1.69 & 1.6388   & H$\alpha$  & $1.39\pm 0.03   $& $8.74_{-0.03}^{+0.04}$ &  $3.37\pm 0.15 $& $11.0_{-1.3}^{+1.0}$  & 1.01\\
AS1063-1904 & $342.1546074 \, -44.5225565$ & 1.73 & 2.0659  & [OIII]\tablefootmark{f}   & $1.55 \pm 0.02  $& $10.11_{-0.01}^{+0.03}$ & - & $6.6_{-0.1}^{+0.5}$ & $-0.69$\\
AS1063-606\_1 & $342.1707948 \, -44.5491248$ & 2.62  & 2.3015 & H$\alpha$,[OIII]\tablefootmark{f},H$\beta$ & $2.13 \pm 0.03$ & $9.65_{-0.05}^{+0.05}$ & $ 3.52 \pm 0.29 $&$19.0_{-4.7}^{+6.5} $& 0.17 \\
AS1063-942 &  $342.1610461 \, -44.5381665$ & 2.21  &2.5803 &  [OIII]\tablefootmark{f},H$\beta$    & $4.75 \pm 0.17  $& $7.42_{-0.08}^{+0.09}$ &  -  & $0.48_{-0.04}^{+0.13}$ & 0.94 \\
AS1063-885\tablefootmark{e} &    $342.1731519 \, -44.5399842$ & 2.00 & 3.1171   & [OIII] & $11.2 \pm 0.46$ &$ 7.21_{-0.01}^{+0.01}$ & - & $0.34_{-0.01}^{+0.01}$ & 0.97\\
AS1063-1720 & $342.2053289 \, -44.5154793$ &  2.84 & 3.5457  & [OIII]  & $3.56 \pm 0.12$ & $10.36_{-0.09}^{+0.07}$ & - & $10.5_{-1.4}^{+3.7}$ & $-0.90$\\

MACS1206-2005 & $181.5630588 \, -8.7869808$ & 1.62 & 1.4295 & H$\alpha$   & $1.53 \pm 0.08$ & $9.89_{-0.03}^{+0.03}$ & $8.23\pm 0.49 $& $51.5_{-9.3}^{+4.5}$	&  0.56     \\
MACS1206-1123 & $181.5690990 \, -8.8022439$ & 1.64 & 1.6421 &  H$\alpha$  & $3.21 \pm 0.04$ & $8.92_{-0.05}^{+0.07} $&$ 1.41\pm 0.12$& $5.2_{-0.8}^{+1.4}$ &   0.49    \\
MACS1206-1409\tablefootmark{e} & $181.5715377 \, -8.7977022$ & 1.68 & 1.7112 & H$\alpha$  &  $2.07 \pm 0.07$    & $9.79_{-0.03}^{+0.03}$  &$ 1.68\pm 1.1$ & $17.0_{-4.1}^{+1.4}$  &  0.10\\
MACS1206-1235 & $181.5266014 \, -8.8002829$ & 1.85 &1.7217 &  H$\alpha$  & $1.83 \pm 0.08$ & $9.10_{-0.02}^{+0.03}$ & $6.58\pm 2.4$ & $ 24.7_{-1.5}^{+1.9}$   & 0.96    \\
MACS1206-445\tablefootmark{e} & $181.5483326 \, -8.8180914$ & 2.08 &1.9125  &  [OIII] &   $1.63 \pm 0.08$  & $9.30_{-0.01}^{+0.03}$  & - & $6.0_{-0.4}^{+0.4} $& $-0.09$    \\
MACS1206-415 & $181.5540750 \, -8.8189422$ &  2.20 & 2.0710 &  H$\alpha$,[OIII]  & $1.62 \pm 0.08$ & $10.02_{-0.06}^{+0.05}$ & $3.03\pm0.72  $& $8.4_{-0.1}^{+0.1}$  &    $-0.51$\\
MACS1206-1622 & $181.5653243 \, -8.7935456$ & 2.44 & 2.0967    &  H$\alpha$  & $ 2.26 \pm 0.15$ & $9.86_{-0.06}^{+0.01}$ & $2.57\pm 0.25 $& $36.9_{-2.5}^{+10.7}$  & $0.29$\\
MACS1206-2372 & $181.5552845 \, -8.7885902$ & 1.97 & 2.3409 & H$\alpha$   & $2.12 \pm 0.16$ & $9.54_{-0.04}^{0.03}$ & $4.64\pm 0.69 $& $10.5_{-0.8}^{+1.0} $& $ 0.02$\\
MACS1206-1616\_1 & $181.5704100 \,-8.7944666$ & 2.86 & 2.4145 & H$\alpha$   & $2.01 \pm 0.11$ & $9.86_{-0.01}^{+0.03}$ & $4.69\pm 0.98 $ & $58.0_{-5.4}^{+3.9}$ & 0.42\\
MACS1206-1472\tablefootmark{d} & $181.5626048 \, -8.7967459$ & 2.43 & 3.0380  & [OIII]  & $4.92 \pm 0.47$ & $9.51_{-0.09}^{+0.01}$ & - & $5.6_{-0.6}^{+0.6}$  & $ -0.30$\\
MACS1206-820\tablefootmark{d} &  $181.5605720 \, -8.8089877$ & 3.04 & 3.0380  & [OIII] &  $7.90 \pm 0.56$   & $8.64_{-0.01}^{+0.06}$ & - & $9.0_{-0.3}^{+0.9}$& 0.84\\

\noalign{\smallskip}
\hline                                   
\end{tabular}
     \tablefoot{\\
     \tablefoottext{a}{The magnifications $\mu$ from the mass models of \citet{Richard2014} and \citet{Cava2017} have been used for AS1063 and MACS1206-08, respectively. 
     }\\
      \tablefoottext{b}{ M$_\star$, SFR$_{SED}$ and SFR$_{H\alpha}$ are corrected for \citet{Chabrier2003} and for the magnification. SFR$_{H\alpha}$ is not corrected for the dust attenuation.}\\
            \tablefoottext{c} {$\Delta$SFR is defined as $\rm \Delta SFR = log(SFR/SFR_{MS})$. We use the main sequence relation from \citet{Tomczak2016} to determine $\rm SFR_{MS}$.
        }\\
        \tablefoottext{d} {Multiple images of the same galaxies. MACS1206-820 is an image of only the top left clump of MACS1206-1472 (see Fig. \ref{image3} in Appendix \ref{appA}).}\\
      \tablefoottext{e} {These three galaxies are unresolved. }\\
      \tablefoottext{f} {The [OIII] in this table refers to the [OIII]$\lambda$5007 emission line, except in these three cases where both [OIII]$\lambda\lambda4960,5007$ are detected. }
      }
\end{sidewaystable*}

%__________________________________________________________________

\subsection{Integrated properties}

\subsubsection{Line fitting}

We first create an integrated spectrum for all the galaxies by summing the spaxels to maximize the signal-to-noise. We fit 1000 Gaussian curves using a Monte Carlo technique perturbing the flux on each of the detected emission lines. The continuum level is determined by the average of all the pixels in a window of $50 \, \AA$ around the center of the emission line. In this window, all the pixels affected by skylines are masked. When the window is populated by too many skylines, we extend it to $70 \, \AA$. If the line is on the edge of the band, the window is cut before the edge and extended on the other side. As a result, we obtain an accurate systemic redshift, the H$\alpha$ and/or [OIII] flux, the full width half maximum (FWHM), the velocity dispersion for each galaxy and their associated uncertainties.

The H$\alpha$ flux is measured for 17 galaxies in our sample.We use it to derive the star formation rate, SFR$_{H\alpha}$, following the \cite{Kennicutt1998} equation~: 

\begin{ceqn}
\begin{align}
\mathrm{SFR}_{H\alpha} = 7.9 \times 10^{-42} \; L(H\alpha) \times \frac{1}{1.7} \times \frac{1}{\mu}
\end{align}
\end{ceqn}
where the factor 1.7 is the correction for the \citet{Chabrier2003} IMF, and $\mu$ is the magnification. Individual magnifications have been derived using the most up-to-date mass models of the clusters. The magnifications for AS1063 were derived by the CATS team as part of the Frontier Fields challenge \citep{Richard2014}. For MACS1206-08, we use the model described in \citet{Cava2017} constrained by many different regions of a giant arc in the cluster core. We have compared that the magnification values derived were similar to the ones previously derived by \citet{Zitrin2009,Zitrin2013}, and obtained through the Hubble Space Telescope Archive, as a high-end science product of the CLASH program \citep{Postman2012}.
We obtain a mean and median fractional difference of 1.4 and 1.06, respectively for AS1063 and MACS1206-08, with a spread of 0.06, which means that our magnification values are slightly higher. 
Only the SFR and stellar mass are affected by the magnification.

We derive the integrated velocity dispersion, $\sigma_{int}$, by correcting the observed velocity dispersion, $\sigma_{obs}$, for the instrumental broadening, $\sigma_{instr}$, which is obtained from the skylines :

\begin{ceqn}
\begin{align}
    \sigma_{int} = \sqrt{ \sigma_{obs}^2 - \sigma_{instr}^2}
\end{align}
\end{ceqn}
These values are listed in Table \ref{table_kinematics1}. The median and average $\sigma_{int}$ of our KLENS sources are $\rm59\, km \, s^{-1}$ and $\rm 64\, km \, s^{-1}$, respectively.

\subsubsection{SED fitting}

We adopted a modified version of the code {Hyperz} (\citeauthor{Bolzonella2000} \citeyear{Bolzonella2000}; \citeauthor{Schaerer2010} \citeyear{Schaerer2010}) to perform the SED fit of the photometric data of our sources. In practice we use the 15 HST bands from the CLASH survey reaching from the bluest (WFC3 UVIS F225W) to the reddest filter (WFC3 IR F160W) plus the IRAC photometry at 3.6 and 4.5 $\mu$m. The photometry was taken from the {RAINBOW} database (cf.\ Sect.\ \ref{section2.1}), relying on a recent multiwavelength 
catalog realised by Rodriguez-Mu\~noz et al.\ (in preparation).
For the SED fits we fixed the redshift to the  spectroscopic value. We have adopted \citet{Bruzual2003} stellar tracks at solar metallicity, close to the values expected for most of our galaxies, and for comparison with other studies. We have assumed exponentially declining star formation histories with timescales $\tau \ge 300$ Myr, including constant SFR. The age is  a free parameter in the fits,  assuming $t>50$~Myr to avoid unphysically young solutions. Nebular emission has been neglected in our default models, as it was found not to affect significantly the resulting stellar masses and to allow meaningful comparisons with other analysis, which also neglect emission. The attenuation is described by the  \cite{Calzetti2000} law and is varied from $A_V=0$ to a maximum value $A^{\rm max}_V=2$. Since model assumptions made here are very similar to those of other studies \citep{ForsterSchreiber2009,Swinbank2017} used for comparison in this paper, they should allow meaningful relative comparisons between these parameters.

The main physical parameters of interest here, stellar mass and the current SFR, are derived from the best-fit SEDs. All results are rescaled to the \cite{Chabrier2003} IMF. To estimate the uncertainties we carry out Monte Carlo simulations perturbing the photometry of the sources.
Fitting 1000 realisations of each source allows us to determine the probability distribution function 
of these parameters. The corresponding median values and 68\% confidence range, corrected for gravitational
magnification, are listed in Table \ref{table_integrated_properties} and presented in Fig. \ref{hist_klens} (middle and right panels). Adopting lower metallicities could lead to somewhat higher masses \citep[see e.g.][]{Yabe2009}. The largest uncertainty in the SFR is the assumption of the star formation history and age prior \citep{Schaerer2013, Sklias2014}. Again, the assumed star formation histories are very similar to those of other studies \citep{ForsterSchreiber2009,Swinbank2017} used for comparison here.

\begin{figure}
   \centering
   \includegraphics[scale=0.38]{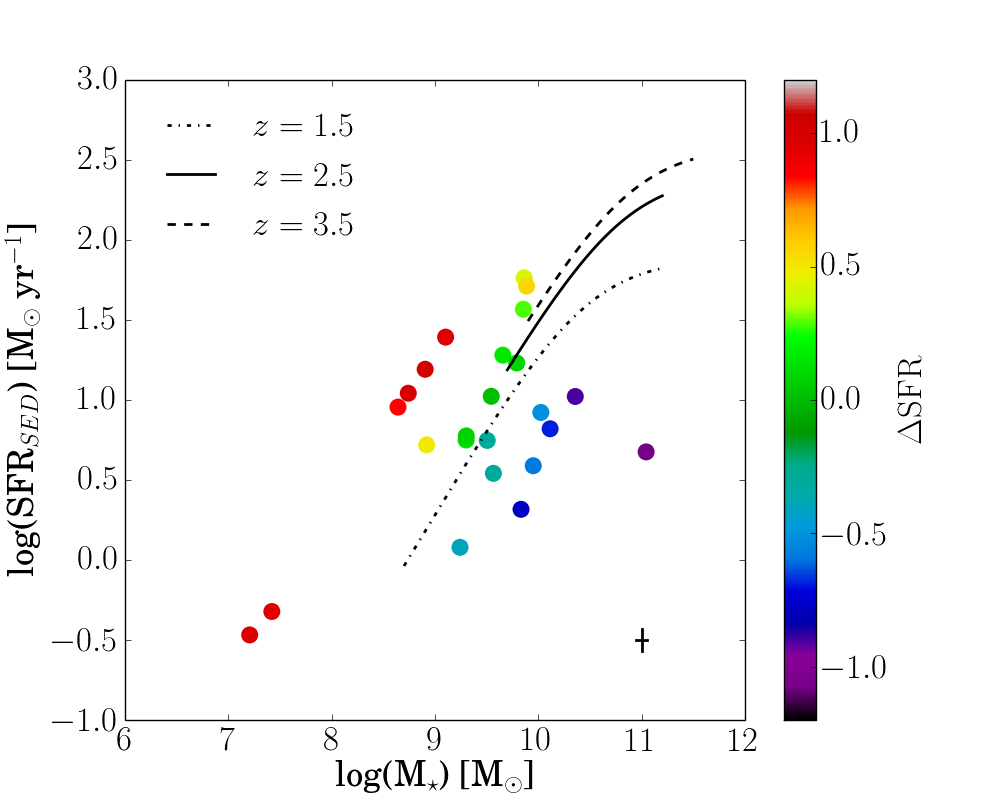}
   \caption{Star formation rate from the SED fitting as a function of stellar mass for all the galaxies in our sample. The curves represent the main sequence of galaxies derived by \citet{Tomczak2016} for the redshifts of 1.5, 2.5, and 3.5. The color corresponds to the value of  $\rm \Delta SFR = log(SFR/SFR_{MS})$, the offset from the main sequence. The typical error bar for individual points is shown in the lower right corner. }
              \label{main_seq}%
   \end{figure}

Figure \ref{main_seq} presents the distribution of galaxies in our sample in the diagram of SFR$_{SED}$ as a function of stellar mass, color-coded as a function of $\rm \Delta SFR$, which is defined as $\rm \Delta SFR = log(SFR/SFR_{MS})$, namely the offset from the main sequence (MS) derived by \citet{Tomczak2016} for the redshift of each galaxy.

%__________________________________________________________________
 %Figure MAIN SEQUENCE

%__________________________________________________________________

%%%%%%%%%%%%%%%%%%%%%%%%%%%%%%%%%%%%%%%%
%           Section4
%%%%%%%%%%%%%%%%%%%%%%%%%%%%%%%%%%%%%%%%

\section{Morphology}
\label{section4}

To further characterize the galaxies in our sample, an analysis of their morphology is required from high-resolution images. The morphology can give us information about the light profile and size as well as helping to distinguish galaxies in interaction, in an on-going merger, or with a clumpy disc. A visual inspection of the HST images shows that our sample includes a large variety of objects, such as compact galaxies, clumpy galaxies, and disc-like galaxies (see Fig. \ref{image1}-\ref{image3} in Appendix \ref{appA}). However, it is impossible to resolve any smaller structure such as spiral arms or a bar.
We use the HST images in the image plane since differential amplification is negligible in our galaxies. Indeed, we obtain a median relative differential amplification of 6\% with a spread of 2\% within our objects. The median absolute differential amplification is $\sim0.12$, which is of the same order as the median error on the amplification of $\sim 0.1$.

To classify our galaxies in a systematic way, we use \mbox{GALFIT} to perform 2D fitting of light profile on 2D images \citep{Peng2002}. We focus on the available HST/WFC3 F160W near-infrared images, which trace the rest-frame optical to the near-UV depending on the redshift of the galaxy, to extract a single image for each galaxy. We take the largest region possible around the galaxy that is not contaminated by light from other galaxies. We adopt a similar approach as \citet{vanderWel2012} and \citet{Wisnioski2015} by determining a single S\'{e}rsic model per galaxy. To obtain a better fit, \cite{Rodrigues2017} use a more complex approach with a S\'{e}rsic model for the bulge and an exponential profile for the disc. However, our sample includes a large diversity of morphology to which a single-S\'{e}rsic fit is most appropriate. 
 
All parameters are free to vary during the fitting process. We obtain as a result  morphological parameters such as the effective radius ($R_e$), the position angle ($PA_{morph}$), the axis ratio ($b/a$), and the S\'{e}rsic index ($n$) for each galaxy. The S\'{e}rsic index, which is an indication of the light profile, approaches in most cases an exponential profile rather than a Devaucouleurs profile \citep{Sersic1963}.
The derived $R_e$ are discussed in Sect. \ref{section51} and Fig. \ref{hist_re}.

%-------------------------------------------------------------
%TABLE KINEMATICS
\begin{table}
\caption{Kinematic properties from fitted models}             
\label{table_model}      
\centering          
\begin{tabular}{l  c c  c c c }     
\hline\hline       
Objects  &  $\upsilon_{rot}$\tablefootmark{a} & $\sigma_{0,galpak}$ & $\rm \Delta PA$\tablefootmark{b} %&  Class
\\
  &  [$\rm km \, s^{-1}$]& [$\rm km \, s^{-1}$]& [$^{\circ}$] &    
\\ 
\hline   
   AS1063-658 &   $120 ^{+52}_{-45}$   & - & $13 \pm 6 $  \\
    AS1063-1670 & $ 217\pm21$  & $38\pm7$ & $ 29 \pm 7$ \\
    AS1063-1152 &   $72\pm7$ & $40 \pm3$ & $6 \pm 5$\\  
   AS1063-1321 &  $ 60 \pm 3$ & $32 \pm 2$ & $5 \pm 5$ \\
   AS1063-732 &    $189^{+27}_{-35}$   & -  &  - \\
   AS1063-1904 &  $42^{+23}_{-25}$ & - & $ 34\pm 8$\\ 
    AS1063-606\_1 &  $50^{+17}_{-28}$ & - & $10 \pm6$  \\  
    AS1063-942 &   $39\pm 16$   & $44 \pm 2$ & $10 \pm 6$ \\
   AS1063-1720 &  $80^{+42}_{-26}$ & - & $50 \pm 7$\\ 
  
     MACS1206-2005 & $184 \pm 41$ &  $83 \pm3$ & $9 \pm 5$ \\
    MACS1206-1616\_1 &  $151^{+96}_{-98}$ &  -& $84\pm7$ \\
 %  MACS1206-2372 & 99 & - \\
   MACS1206-1472 & $172 \pm 92  $& - & $5\pm7$ \\

\noalign{\smallskip}
\hline                                   %inserts single line
\end{tabular}
     \tablefoot{\\
     \tablefoottext{a}{ $\upsilon_{rot}$ is corrected for the inclination.
     }\\
     \tablefoottext{b}{$\rm \Delta PA$ is defined as $\lvert PA_{kin} - PA_{morph} \lvert$. No value is obtained for $PA_{morph}$ in the case of AS1063-732 due to the clumpy morphology. } }
\end{table}

%-------------------------------------------------------------

%------------------------------------------------------
%TABLE KINEMATICS
\begin{table*}
\caption{Kinematic properties}             
\label{table_kinematics1}      
\centering          
\begin{tabular}{l  c c c c c c c  c c }      
\hline\hline       
Objects  & $\sigma_{int}\tablefootmark{a}$ & $\sigma_{0,lim}$\tablefootmark{b} & $\upsilon_{obs,lim}$\tablefootmark{c} & $\upsilon_{obs,lim}/\sigma_{int}$ & $\upsilon_{obs,lim}/\sigma_{0,lim}$ 
\\
   & [$\rm km \, s^{-1}$] & [$\rm km \, s^{-1}$]  & [$\rm km \, s^{-1}$]  &  &   
\\ 
\hline   

   AS1063-1816 &  44  $\pm \, 3$   &  - &  - & - & - & %3 & 3    
   \\
   AS1063-658 &   62  $\pm \, 4$    &  $32 \pm 9 $& $63 \pm 10 $ & 1.01      & 1.97    %&1 & 2  
   \\
   AS1063-1611 &  48   $\pm \, 2$  &  - &  - & - & - &% 3 & 3      
   \\
   AS1063-1670 &  119  $\pm \, 27$  & $32  \pm 12$ & $80 \pm 10 $& 0.67 & 2.5  & %1 & 1 
   \\
  AS1063-1152 &  53 $\pm \, 4$  & $40 \pm 8 $ & $ 80 \pm 11 $ & 1.51     &2.0    & %1 & 1      
  \\  
   AS1063-1321 &  52 $\pm \, 5$  &  $35 \pm 7 $  & $65 \pm 8 $   & 1.25    & 1.86  & %1& 1 
   \\
   AS1063-1331 &  60  $\pm \, 5$  &  - &  - & - & - & %3 & 3        
   \\
   AS1063-732 &   79   $\pm \, 6$ &  $50 \pm 10$ &  $81 \pm 15$  & 1.03      & 1.62  & %1 & 1     
   \\
   AS1063-1904 &  55  $\pm \, 4$     & $66 \pm 9$   & $22 \pm 5$   & 0.39 & 0.33 & %2 & 2 
   \\ 
    AS1063-606\_1 &  40 $ \pm \, 5$  & $  31 \pm 6 $  & $12 \pm 5$  & 0.3      & 0.39  &% 2 & 2   
    \\  
      AS1063-942 &  40 $\pm \, 4$  &   $32  \pm 6  $& $31 \pm 8 $ & 0.78      & 0.97  & %1 & 2       
      \\
    AS1063-885\tablefootmark{d} &   31 $\pm \, 11$  &  - &  - & - & - & %- & -  
    \\
    AS1063-1720 &  105  $\pm \, 54$  & $52 \pm 25$   & $37 \pm 10$    & 0.35 & 0.71  &%2 & 2 
    \\ 

   MACS1206-2005 & 104 $\pm \, 8$ & $82 \pm 15$ & $ 90 \pm 7 $& 0.86 & 1.1 & %1& 1 
   \\
   MACS1206-1123 & 35 $\pm \, 4$ &  - &  - & - & - & %3 & 3   
   \\
   MACS1206-1409\tablefootmark{d} & 40 $\pm \, 22$&  - &  - & - & - & %- & -  
   \\
   MACS1206-1235 & 59 $\pm \, 29$ &  - &  - & - & - & %3 & 3 
   \\
   MACS1206-445\tablefootmark{d} & 71 $\pm \, 15$ &  - &  - & - & - & %- & - 
   \\
   MACS1206-415 & 78 $\pm \, 20$&  - &  - & - & - & %3 & 3 
   \\
   MACS1206-1622 & 48 $\pm \, 5$  &  - &  - & - & - & %3 & 3  
   \\
   MACS1206-2372 & 58 $\pm \, 29$&  -  &  -& -& - & %3& 3 
   \\
   MACS1206-1616\_1 & 77 $\pm \, 20$ & $79 \pm 24$ &   $81 \pm 7$ & 1.05 & 1.02 & %1 & 1
   \\
   MACS1206-1472 & 82 $\pm \, 7$ & $62 \pm 13$  &$ 63  \pm 6 $ & 0.77 & 1.02 & %1 & 2 
   \\
   MACS1206-820 &  96 $\pm \, 23$ &  - &  - & - & - & %3 & 3 
   \\

\noalign{\smallskip}
\hline                                   %inserts single line
\end{tabular}
     \tablefoot{\\
     \tablefoottext{a}{The integrated dispersion $\sigma_{int}$ is the dispersion measured from the integrated spectrum corrected for instrumental broadening (see Eq. 2).}\\
     \tablefoottext{b}{The upper limit dispersion $\sigma_{0,lim}$ is the dispersion measured in the outer region on the major axis to avoid beam smearing effects and corrected for instrumental broadening.} \\
     \tablefoottext{c} {The lower limit velocity $\upsilon_{obs,lim}$ is defined as the difference between the maximum and minimum velocity on the major axis divided by two, not corrected for inclination. When both H$\alpha$ and [OIII] emission lines are detected for the same galaxy, we use H$\alpha$ only to determine $\upsilon_{obs,lim}$.} \\
     \tablefoottext{d} {These three galaxies are unresolved.
     }
     \\
   }
\end{table*}
%
%------------------------------------------------------

%%%%%%%%%%%%%%%%%%%%%%%%%%%%%%%%%%%%%%%%
%           Section5
%%%%%%%%%%%%%%%%%%%%%%%%%%%%%%%%%%%%%%%%

\section{Kinematics}
\label{section5}

\subsection{Modeling}
\label{section51}

We perform Gaussian fits to the H$\alpha$ or [OIII] emission line in individual spaxels to obtain the flux, velocity, and velocity dispersion maps. The results are shown in Fig. \ref{image1}-\ref{image3} in Appendix~\ref{appA}. The spectroscopic redshift obtained from the integrated spectrum has been used to determine the central wavelength for the velocity maps. We use again a spectral window of $50 \, \AA$ to determine the continuum and we mask all the skylines. We also determine the level of noise for each spaxel in the same spectral window. We bin spaxels $2 \times 2$ (0.4" $\times$ 0.4") for the area where the noise is too high, but in most cases the new area is still too noisy (S/N<2). We reject the spaxels where the signal-to-noise ratio is lower than 2. We apply this method for all the galaxies resolved in our sample. The shear introduced by weak-lensing does not affect the kinematics since these low-magnification galaxies are seeing-limited.

To obtain the kinematics of our galaxies, we use two different approaches. First, we perfom fits with the GalPaK$^{\rm 3D}$ code which directly fits a 3D galaxy disc kinematic model to the 3D datacubes (\citeauthor{Bouche2015} \citeyear{Bouche2015}). This code has been used before on datacubes from VLT/MUSE (e.g. \citeauthor{Contini2016} \citeyear{Contini2016}), VLT/SINFONI (e.g. \citeauthor{Schroetter2015} \citeyear{Schroetter2015}), and VLT/KMOS \citep{Mason2017}. The 3D model is convolved with the point spread function (PSF) and the line-spread function (LSF) to derive kinematic properties such as the intrinsic velocity dispersion and rotation velocity, but also morphological properties like the inclination. The PSF in our case has been determined by the seeing obtained during the observations of stars from the acquisition image and the LSF has been measured from the skylines. 

GalPaK$^{\rm 3D}$ is currently configured for an exponential, gaussian, or De Vaucouleurs radial flux profile. We use an exponential profile which is more adapted for our galaxies.

We adopt the arctangent function for the velocity profile \citep{Courteau1997}, which has been used in many studies (e.g. \citeauthor{Jones2010} \citeyear{Jones2010}):

\begin{ceqn}
\begin{align}
\label{v_rot}
    \upsilon(r) = \upsilon_{rot} \frac{2}{\pi} \, \mathrm{arctan} \frac{r}{r_t}
\end{align}
\end{ceqn}
where r is the radius, $r_t$ is the turnover radius, and $\upsilon_{rot}$ is the maximum rotation velocity. 

GalPaK$^{\rm 3D}$ has 10 parameters : the position of the center $(x,y,\lambda)$, the disc half-light radius, $R_{1/2}$, the flux, the inclination, the position angle for the kinematics, $PA_{kin}$, the turnover radius, $r_t$, the maximum rotation velocity, $\upsilon_{rot}$, and the intrinsic dispersion, $\sigma_{0 \, galpak}$. We constrain the position centre $(x,y)$ in 2 spaxels according to the flux map and $\lambda$ in a window of twice the FWHM found with the integrated spectrum around the center of the emission line. The other parameters are free to vary. Since the number of spaxels with good signal-to-noise in our galaxies is small, we have been  able to fit only five galaxies with GalPaK$^{\rm 3D}$.

For the other galaxies showing a velocity gradient in their velocity map but that are overfitted with GalPaK$^{\rm 3D}$, we obtain a 2D disc model following the arctangent function (Eq. \ref{v_rot}) using the MCMC method from \citet{Leethochawalit2016}. This approach is used to fit the velocity map with the PSF convolved model. The spatial center is beforehand constrained to less than 2 spaxels and 
the inclination, the position angle for the kinematics, $PA_{kin}$, the turnover radius, $r_t$, and the maximum rotation velocity, $\upsilon_{rot}$, are free to vary. The results of the 12 galaxies for which we have obtained model fits are shown in Fig. \ref{image4} in Appendix~\ref{appA} and are given in Table \ref{table_model} with the velocity dispersion found by GalPaK$\rm^{3D}$ when the kinematic fit with this tool converged.

%__________________________________________________________________

  %Figure R/RE
   \begin{figure}
   \centering
   \includegraphics[scale=0.37]{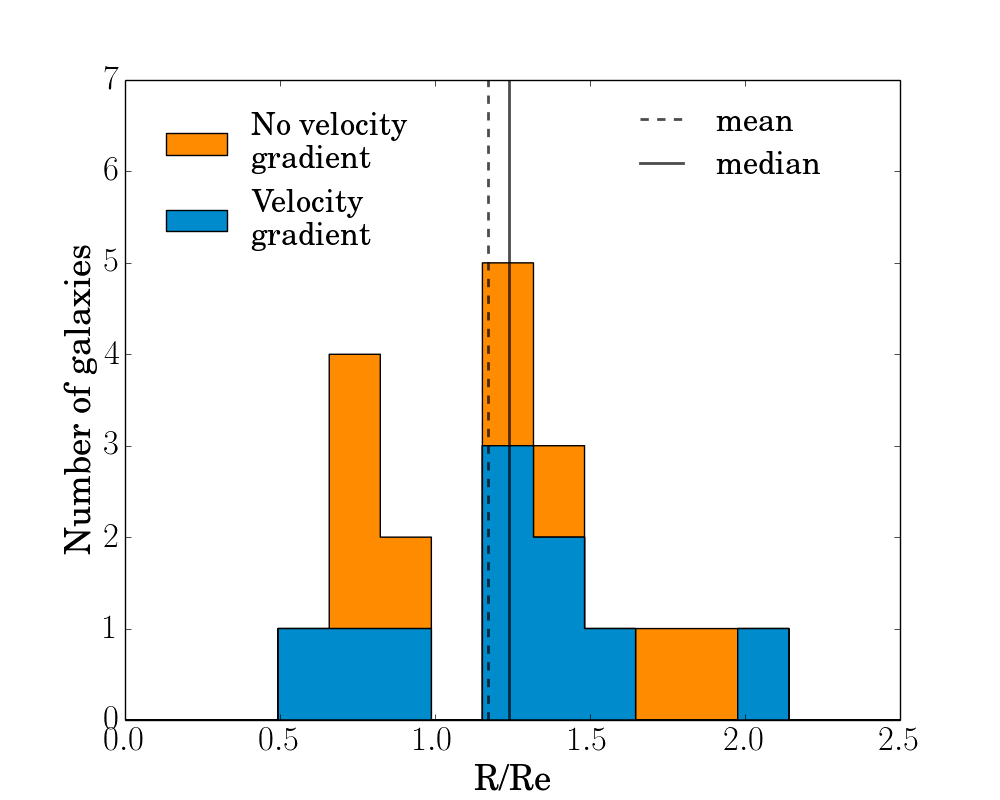}
   \caption{Histogram of $R$/$R_e$ for the resolved galaxies of our sample, where $R$ is the maximum radius where we get a measurement of the velocity and $R_e$ is the effective radius. Only galaxies for which it is possible to fit a light profile with \mbox{GALFIT} are shown. 
   The galaxies that do not show a velocity gradient are in orange and the galaxies for which it is possible to fit a kinematic model (velocity gradient) are presented in blue. 
   }
              \label{hist_re}%
   \end{figure} 

%__________________________________________________________________

The median value of $R$/$R_e$ in our sample is 1.2, where $R$ is the maximum radius where we get a measurement of the velocity, and $R_e$ is the effective radius derived from 2D GALFIT fitting (Sect. \ref{section4}). The histogram of $R$/$R_e$ for the galaxies in our sample is plotted in Fig. \ref{hist_re}. \cite{Genzel2017} show that the rotation velocity, $\upsilon_{rot}$, reaches a maximum between $ 1.3 \,R_e$ and $ 1.5 \,R_e$, whereas in \citet{Lang2017} they obtain $ 1.65 \, R_e$ by stacking more than 100 galaxies. \citet{Stott2016} are taking measurements of the rotation velocity at $ 2.2~R_e$, and \citet{Turner2017}  at $ 2.0 \,R_e$, where the rotation curve already flattens. Therefore, a good signal-to-noise at at least $\sim~1.5  \, R_e$ is needed to obtain an accurate  rotation velocity.

In our sample, we obtain a radius $R$ lower than the $\sim 1.5 \, R_e$ needed for many galaxies (see Fig. \ref{hist_re}). As a result, there is a degeneracy between the different parameters in the kinematic model, and therefore our kinematic model results show large errors. In fact, we get a mean ratio of the rotation velocity from the models, $\upsilon_{rot}$, and the maximum velocity observed from the rotation curve as large as $\sim29\%$, since the modeled rotation velocities are taken from the extrapolation of the rotation curve at large radii. Consequently, in what follows, we prefer to introduce a kinematic classification which does not refer to the uncertain $\upsilon_{rot}$.

%-------------------------------------------------------------
%TABLE CLASSIFICATION
\begin{table*}
\caption{Kinematic classification}             
\label{table_kinematics2}      
\centering          
\begin{tabular}{l  c c c c c c c  c c c }     % 7 columns 
\hline\hline       
                      % To combine 4 columns into a single one 
Objects  & Velocity & $\upsilon_{rot}$  & $\upsilon_{obs,lim}/\sigma_{int}$ &
$\upsilon_{obs,lim}/\sigma_{0,lim} $ & Peak  & $\Delta$PA$\tablefootmark{d}< 30 ^{\circ}$  & No & Class  & Final 
\\ 
 & gradient\tablefootmark{a} & achieved\tablefootmark{b} & $> 0.4$ & $> 1 $ & of $\sigma$\tablefootmark{c} &  & int.\tablefootmark{e} & FS\tablefootmark{f} & class\tablefootmark{g}
\\
\hline   

   AS1063-1816 &        &   &   &       &       &      &  X  & 3&3        \\
   AS1063-658 &   X    &   & X  & X     &       & X     &    &1&2       \\
   AS1063-1611 &       &   &   &        &       &      &  X     &3&3         \\
   AS1063-1670 &  X    &   & X & X      &  X    & X      & X   &1&1         \\
  AS1063-1152 &  X    &    & X  & X     &       & X      & X       &1&1           \\  
   AS1063-1321 &  X   &  X & X  & X     & X     & X     &  X    &1&1        \\
   AS1063-1331 &      &     &   &       &       &      &  X  &3&3                 \\
   AS1063-732 &   X    &    & X  & X    &       &      &  X   &1&1              \\
   AS1063-1904 &  X   &     &    &      &       &      &  X       &2&2      \\ 
    AS1063-606\_1 &  X &    &   &       &       &  X     & X   &2&2 \\  
      AS1063-942 &  X  &     & X &      &       &X      &  X   &1&2    \\
    AS1063-885 &      &   &     &       &       &      & X &  4&4 \\
    AS1063-1720 &  X &    &     &       &       &      & X  &2&2  \\ 

   MACS1206-2005 & X &   &  X & X       &       & X     &  X     &   1&1\\
   MACS1206-1123 &  &   &    &          &       &      & X  & 3&3\\
   MACS1206-1409 &  &   &    &          &       &      & X  & 4&4   \\
   MACS1206-1235 &  &   &    &          &       &      & X  & 3&3 \\
   MACS1206-445 &   &   &    &          &       &      &  X & 4&4 \\
   MACS1206-415 &  &   &     &          &       &      &  X & 3&3 \\
   MACS1206-1622 &  &   &     &         &       &      &  X  & 3&3  \\
   MACS1206-2372 &  &    &   &          &       &      & X & 3&3\\
   MACS1206-1616\_1 & X & & X & X       &       &      & X & 1&1\\
   MACS1206-1472 & X &   & X & X        &       & X     & & 1&2\\
   MACS1206-820 &   &   &    &          &       &      &   & 3&3\\

\noalign{\smallskip}
\hline                                   %inserts single line
\end{tabular}
     \tablefoot{The cross (X) indicates that the criterion is satisfied.
     \\
       \tablefoottext{a}{A velocity gradient is visible in the velocity map of the galaxy.}\\
         \tablefoottext{b}{ $\upsilon_{rot}$ achieved when a flat part is visible in the rotation curve of the observed velocity.}\\
     \tablefoottext{c}{A peak in the dispersion is observed in the dispersion map at the center of the galaxy.}\\
     \tablefoottext{d}{$\Delta$PA is defined as $\lvert PA_{kin} - PA_{morph} \lvert$.}\\
        \tablefoottext{e}{The galaxies classified as no interaction (no int.) do not show any evidence of disturbance in their velocity dispersion map obtained from H$\alpha$ or [OIII] emission lines. A disturbance could be the signature of a system in an on-going merger. } \\
       \tablefoottext{f} {Criteria established by \citet{ForsterSchreiber2009} : galaxies with $\upsilon_{obs,lim}/\sigma_{int} > 0.4$ are defined as rotation dominated (1) and with $\upsilon_{obs,lim}/\sigma_{int} < 0.4$ as dispersion dominated (2). Galaxies classified as (3) do not show any evidence of a velocity gradient. Unresolved galaxies are classified as (4).} \\
     \tablefoottext{g} {The ratio $\upsilon_{obs,lim}/\sigma_{0,lim}$ is the criterion we use in this work. Galaxies with $\upsilon_{obs,lim}/\sigma_{0,lim} > 1$ are classified as rotation dominated (1) if they also show no sign of interactions in their dispersion map (no int.), otherwise they are classified as irregular (2). The galaxies classified as (3) do not show any evidence of a velocity gradient. Unresolved galaxies are classified as (4). }\\
     }
\end{table*}

%----------------------------------------
\subsection{Empirical diagnostics}
\label{section52}

Another way to obtain information about the kinematics is by using the lower limit of the rotation velocity and the upper limit of the intrinsic velocity dispersion. For all the galaxies showing a gradient in the velocity field, we take the maximum and the minimum velocities observed on the major axis to obtain a lower limit on the rotation velocity :

\begin{ceqn}
\begin{align}
    \upsilon_{rot} \geq  \upsilon_{obs,lim} = \frac{\upsilon_{max \, obs} - \upsilon_{min \, obs}}{2}.
\end{align}
\end{ceqn}

This value is not corrected for inclination and $\upsilon_{obs,lim}$ is then a lower limit (Table \ref{table_kinematics1}).

Different methods have been recently used to determine the intrinsic velocity dispersion and to correct for the beam smearing effect. \citet{Johnson2017}, for example, measure the velocity dispersion in the outskirt of the galaxies, but also measure the median of all available spaxels. They afterwards apply beam smearing corrections that they derive from modeling the median values. Several studies use this first method and measure the intrinsic velocity dispersion where the beam smearing is expected to be negligible, namely in the outer regions on the major axis of galaxies \citep[e.g.][]{ForsterSchreiber2009,Wisnioski2015}. Recent tools have also been developed to model and extract the intrinsic galaxy dynamics directly from the datacubes like GalPaK$\rm^{3D}$ \citep{Bouche2015}, and 3D-Barolo \citep{DiTeodoro2015}. Moreover, several studies use the beam smeared local velocity gradient to correct linearly the observed velocity \citep[e.g.][]{Stott2016, Mason2017}.

In our case, the intrinsic velocity dispersion has been determined by using spaxels on the major axis showing the lowest values in the most outer regions of the galaxies to avoid beam smearing effects as much as possible. Beam smearing is most significant for massive galaxies, galaxies with a high inclination, and compact galaxies \citep[e.g.][]{Newman2013, Burkert2016, Johnson2017}. For our sample with a median mass of $\rm log(M_\star/M_\odot)=9.6 $, a median inclination of $\sim45^\circ$ and a median value of $R_e/R_{PSF} \sim 1.5$, we expect that the beam smearing increases our measured velocity dispersions by less than $\sim 10\%$, which is of the same order as the typical error on the measurements.

Since the beam smearing is expected to be negligible in our galaxies,
and to be consistent with the measurements of \citet{ForsterSchreiber2009} and \citet{Wisnioski2015}, two of our largest comparison samples, we do not apply here any extra correction for the beam smearing after measuring the lowest velocity dispersion in the outskirts. We correct the value for instrumental broadening which we measure from the skylines. We therefore obtain an upper limit on the velocity dispersion :

\begin{ceqn}
\begin{align}
    \sigma_{0} \leq  \sigma_{0,lim} = \sqrt{ \sigma_{outer}^2 - \sigma_{instr}^2}.
\end{align}
\end{ceqn}

The derived measurements are available in Table \ref{table_kinematics1}. We note that for the galaxies modeled with Galpak$^{\rm 3D}$, the values obtained for $\sigma_{0,lim} $ are in good agreement with  $\sigma_{0,galpak}$ since we obtain a difference of less than $\sim7\%$, indicating that our estimation of $\sigma_{0}$ with $\sigma_{0,lim}$ is realistic.
\\

%----------------------------------------

\subsection{Classification}

Rotation dominated galaxies have generally been classified using the ratio $\upsilon_{rot}/\sigma_0 > 1$ (e.g \citeauthor{Wisnioski2015} \citeyear{Wisnioski2015}). We also adopt this definition here, but in our case with  $\upsilon_{obs,lim}$ and $\sigma_{0,lim} $, which gives  $\upsilon_{obs,lim}/\sigma_{0,lim} > 1$. Since $\upsilon_{obs,lim}$ is a lower limit and $\sigma_{0,lim}$ is an upper limit, if $\upsilon_{obs,lim}/\sigma_{0,lim}>1$, one also has $\upsilon_{rot}/\sigma_0 >1$ by definition.

We divide our galaxies into four kinematic classes : 

\begin{itemize} 
\item (1) Rotation dominated : All galaxies with $\upsilon_{obs,lim}/\sigma_{0,lim} > 1$ and with no evidence of disturbance in the dispersion map to avoid galaxies in interaction. 

\item (2) Irregular rotators : All galaxies showing a gradient of velocity, but not classified as rotation dominated. The classification is uncertain for these galaxies. Most of them show $\upsilon_{obs,lim}/\sigma_{0,lim} < 1$. Some galaxies show $\upsilon_{obs,lim}/\sigma_{lim} > 1$, but also some signs of disturbance and interactions in their dispersion maps (see Fig. \ref{image1}-\ref{image3}).

\item (3) Non-rotators : No gradient in the velocity map is observed, objects are clumpy, compact, or with an irregular velocity map.

\item (4) Unresolved : Three galaxies in our sample are not resolved.
\end{itemize}

The position angle can also be an important indicator to classify the kinematics of galaxies. \citet{Rodrigues2017}, \citet{Wisnioski2015}, and several other studies suggest a large discrepancy between the position angle obtained from the morphology and kinematics is an indication of clumpy structure or interaction. A difference in  $\lvert PA_{kin} - PA_{morph} \lvert = \Delta \mathrm{PA}$ up to $ 30^{\circ} $ is usually tolerated. We find a mean and median of $23^{\circ}$ and $10^{\circ}$ with a spread of 7.5$^{\circ}$ respectively, meaning most of the galaxies in our sample are aligned (among the 12 objects with measured  $PA_\mathrm{kin}$ values). 
We have only three\footnote{There is also a fourth object, AS1063-732, for which it has been impossible to define $\Delta$PA since the galaxy shows evident clumps in the HST image.} objects where $\Delta$PA is clearly higher than $ 30^{\circ} $. One of these galaxies is classified as rotation dominated and the other two as irregular.

A summary of our kinematic classification for each galaxy in KLENS can be found in Table \ref{table_kinematics2}. We see a velocity gradient in 12 galaxies. The flat part of the rotation curve is reached in only one galaxy and a peak of dispersion is seen in only two galaxies. The classification of \citet{ForsterSchreiber2009}, referred as FS, is based on the criteria $\upsilon_{obs,lim}/\sigma_{int}>0.4$ for rotation dominated galaxies and $\upsilon_{obs,lim}/\sigma_{int}<0.4$ for dispersion dominated galaxies. Finally, the last column gives our final classification.

From our classification, we obtain a rotation dominated galaxy fraction of  $\sim 25 $\% (only resolved galaxies). Using the FS classification, we find a fraction of $\sim40\%$ as rotation dominated and a fraction of $\sim 15\%$ as dispersion dominated. Adding the $\Delta$PA  criterion, the fractions of rotation dominated galaxies become 15\% and 30\% for our classification and FS classification, respectively. Clearly, the choice of the classification has a big impact on the fraction and is important in the comparison with other samples, especially for small samples.

%%%%%%%%%%%%%%%%%%%%%%%%%%%%%%%%%%%%%%%%
%           Section6
%%%%%%%%%%%%%%%%%%%%%%%%%%%%%%%%%%%%%%%%

\section{Discussion}

\label{section6}

\subsection{Fraction of rotation dominated galaxies}

We find a fraction of rotation dominated galaxies of 25-40\% for our galaxies with low mass and/or low SFR. From the literature, we know that a low fraction is also observed in recent surveys studying low masses at high redshift such as \citet{Turner2017} with 35\%, \citet{Leethochawalit2016} with 36\%, and \citet{Gnerucci2011} with 30\%. \citet{Mason2017} have also obtained a fraction of only 12\% when using strict criteria, but the fraction becomes $\sim60\%$ when adding more irregular rotators. \citet{Contini2016} and \citet{Livermore2015} claim that up to 50\% of their analysed galaxies are rotation dominated. Other surveys studying more massive galaxies at $0.8<z<2.5$ obtained a higher fraction. 
\citet{ForsterSchreiber2009}, \citet{Wisnioski2015}, and \citet{Stott2016} find a fraction of $\sim 75$-$80\%$, while \citet{Epinat2012} obtain 65\%. A dependence between the mean redshift of the surveys and the rotation dominated fraction has been observed in recent studies (e.g. \citeauthor{Stott2016} \citeyear{Stott2016}; \citeauthor{Simons2017} \citeyear{Simons2017}; \citeauthor{Turner2017} \citeyear{Turner2017}), but are there
other physical parameters which drive the observed fractions of rotation dominated galaxies?

%--------------------------------------------------------------------------
         \begin{figure}
   \centering
  \subfloat{\includegraphics[scale=0.352]{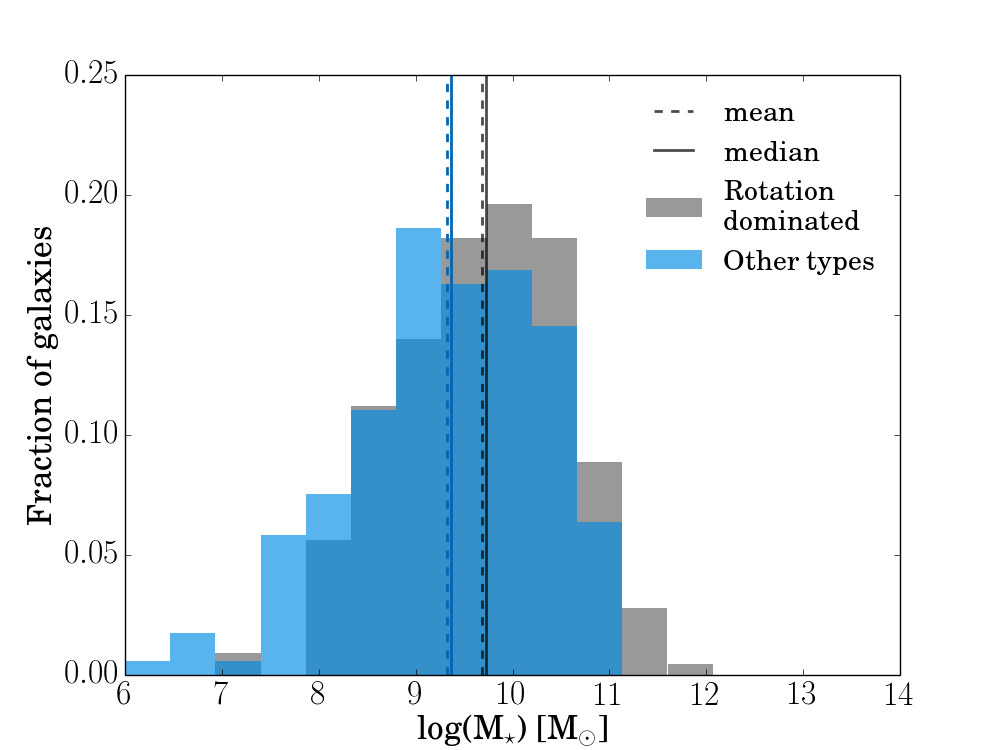}}\\
   \subfloat{\includegraphics[scale=0.352]{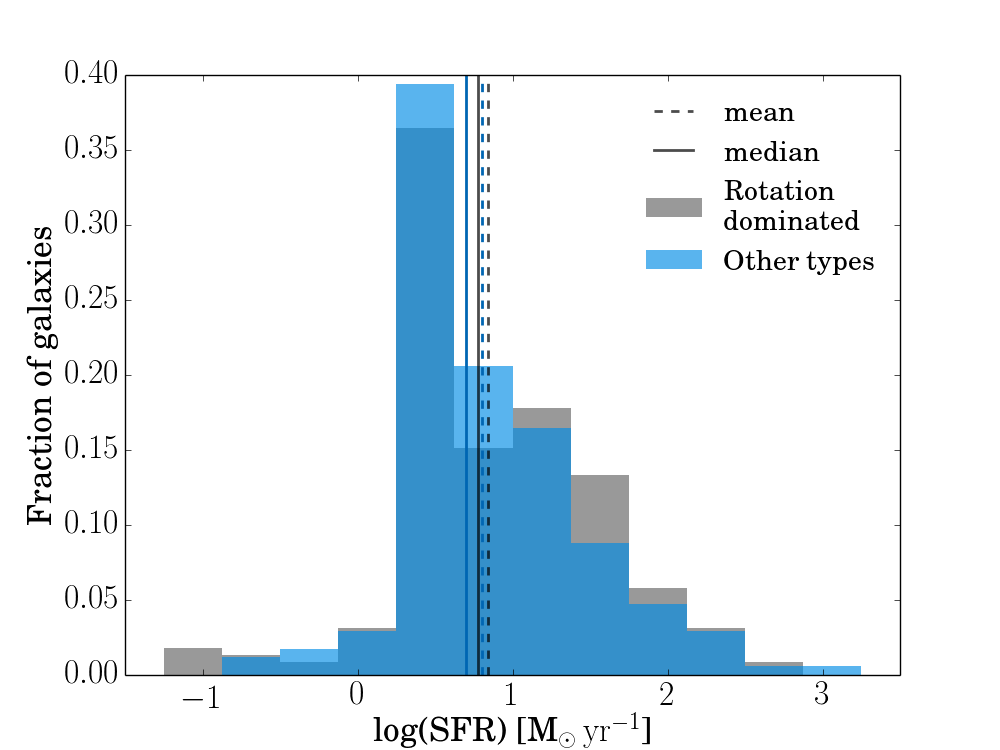}}\\
   \subfloat{\includegraphics[scale=0.352]{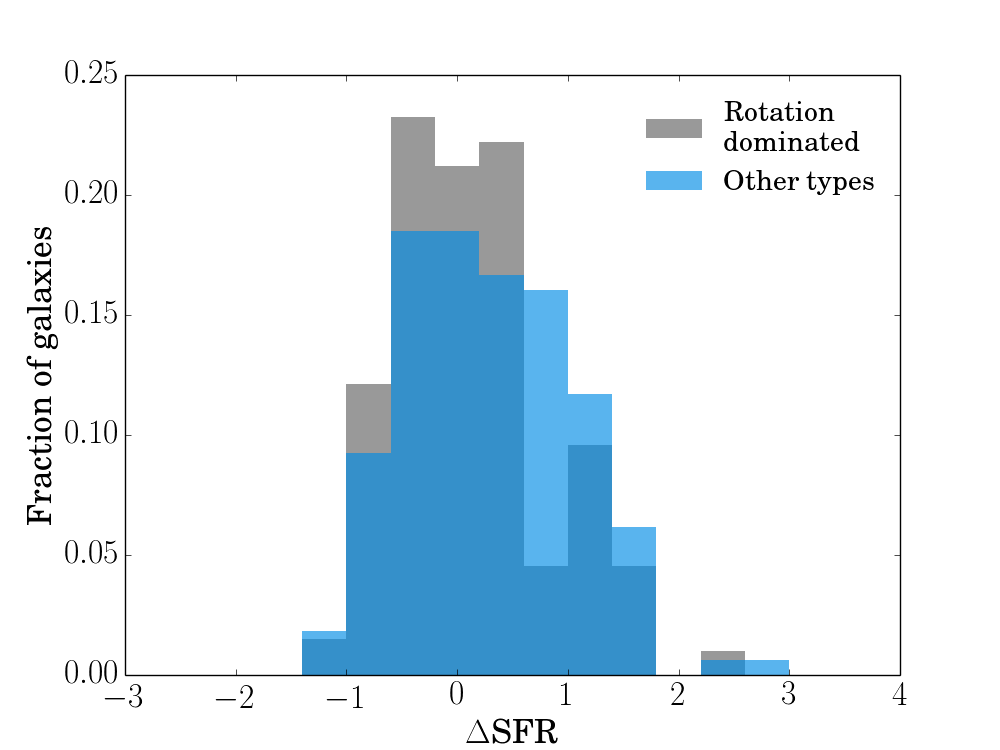}}

      \caption{Histograms of the stellar mass (top panel), SFR (middle panel), and $\Delta$SFR (bottom panel), where $\rm \Delta SFR=log(SFR/SFR_{MS})$, normalized for the rotation dominated galaxies compared to other types. We combine our data with SINS \citep{ForsterSchreiber2009}, \citet{Livermore2015},  MUSE-HDFS \citep{Contini2016}, MUSE-KMOS \citep{Swinbank2017}, KLASS \citep{Mason2017}, and KDS \citep{Turner2017}. KLASS is considered for the mass only, no SFR is available.}
         \label{hist_RD}
   \end{figure}
%--------------------------------------------------------------------------

To investigate this question, we combine our data with the comparison samples listed above (Sect. \ref{comparisonsamples}), which have publicly available information needed to discriminate rotation dominated  galaxies from other types (dispersion dominated, merger, no velocity gradient, etc.).

Figure \ref{hist_RD} presents the respective histograms of the stellar mass, SFR, and $\Delta$SFR (when available) for the galaxies classified as rotation dominated with respect to other types. In the SINS sample the ratio $\upsilon_{obs}/\sigma_{int} >0.4$ is used as a criterion for rotation dominated, while in the other surveys mentioned we classify the galaxies with $\upsilon_{rot}/\sigma_0 >1$ as rotation dominated.

For all the surveys combined, we obtain a median mass of $\rm log(M_{\star}/M_\odot) = 9.73$ with a spread of $ 0.14$ and $9.36 $ with a spread of $ 0.17$ for rotation dominated galaxies and other types, respectively (Fig.~\ref{hist_RD}, top panel). This indicates that massive galaxies are more often predominantly supported by ordered rotation than the low mass galaxies within a significance of 2$\sigma$.

This is in line with what we observe in our sample, the rotation dominated galaxies seem in majority to be the massive ones (see Tables \ref{table_integrated_properties} and \ref{table_kinematics2}) as found by simulations of galaxies in the local Universe \citep{El-Badry2017}. The same trend is visible when we combine the rotation dominated galaxies plus the irregular rotators and compare them with the non-rotators. For our small sample, this trend could also be a bias of the detection limit because the low mass galaxies are less luminous. Therefore, the outer regions are poorly detected, and we possibly do not see the rotation behavior in these galaxies. \citet{ForsterSchreiber2009} have also reported an increase in the fraction of rotators for the massive galaxies of their sample, which was targeting more massive galaxies (median of $\rm log(M_{\star}/M_\odot) \sim 10.4 $) at similar redshifts to our study ($z\sim 1.5-2.5$). \citet{Kassin2012} obtain similar results with the DEEP2 Survey as well as \citet{Simons2016,Simons2017}  for galaxies at $z \sim 2$ with masses of $\rm 10^9-10^{11} \, M_\odot$.

No clear trend is observed for $\rm log(SFR)$ with a median of $0.78 \pm 0.11 $ and $\rm 0.70 \pm 0.11 \, M_\odot \, yr^{-1}$ for rotation dominated and other types, respectively (Fig. \ref{hist_RD}, middle panel). Finally, there is a trend for galaxies showing starburst or intense star formation ($\rm \Delta SFR>0.5$) in Fig. \ref{hist_RD} (bottom panel) to be more likely irregular than the galaxies on the main sequence ($\rm \Delta SFR \sim 0$) or the quenching galaxies ($\rm \Delta SFR<0$).

\subsection{Evolution of the kinematic properties}

In recent studies, it has been pointed out that the velocity dispersion increases with redshift. \citet{Wisnioski2015} have suggested this trend could come mainly from the increase of the gas fraction with redshift, in the framework of the equilibrium model. They propose a model following this idea. The gas fraction can be expressed as :

\begin{ceqn}
\begin{align}
\label{eq1}
f_{gas}(z, \mathrm{M}_{\star})=\frac{1}{1+(t_{depl} \mathrm{sSFR})^{-1}},
\end{align}
\end{ceqn}
\\
where $t_{depl}$ is the molecular gas depletion timescale and sSFR the spectific star formation rate. The  $t_{depl}$ is given by 

\begin{ceqn}
\begin{align}
t_{depl}(z) = 1.5(1+z)^{\alpha} \, \,  \mathrm{[Gyr]},
\end{align}
\end{ceqn}
where the exponent has been fixed at $\alpha =-1$ in the case of \cite{Wisnioski2015} based on the work of \citet{Tacconi2013}. More recently, \citet{Genzel2015} and \citet{Dessauges-Zavadsky2017} obtained a shallower $t_{depl}$ dependance on $z$, which we adopt in this work ($\alpha = -0.85$).

\cite{Wisnioski2015} use the parametrisation of the specific star formation rate determined at $ 0.5 < z < 2.5 $ by \citet{Whitaker2014} :

\begin{ceqn}
\begin{align}
\label{eq2}
\mathrm{sSFR}(z, \mathrm{M}_{\star}) = 10^{a(\mathrm{M}_{\star})}(1+z)^{b(\mathrm{M}_{\star})}  \, \,  \mathrm{[Gyr^{-1}]},
\end{align}
\end{ceqn}
with $ a(\mathrm{M}_{\star})$ and $b(\mathrm{M}_{\star})$ defined as 

$$
a(\mathrm{M}_{\star}) = -10.73 + \frac{1.26}{1+e^{\frac{10.49-\mathrm{M}_{\star}}{-0.25}}}
$$
$$
b(\mathrm{M}_{\star}) = 1.85 + \frac{1.57}{1+e^{\frac{10.35-\mathrm{M}_{\star}}{0.19}}}
$$
\\
where the stellar mass is valid for $\rm 9.2< log( M_\star/M_\odot) <11.2  $.

Finally, following the Toomre equilibrium criterion \citep{Toomre1964}, one gets

\begin{ceqn}
\begin{align}
\label{eq_sigma}
\sigma_0(z, \mathrm{M}_{\star}) = \frac{\upsilon_{rot} \, f_{gas}(z, \mathrm{M}_{\star})  \, Q_{crit}}{a}  \, \,  \mathrm{[km \, s^{-1}]},
\end{align}
\end{ceqn}
where $ a = \sqrt{2}$ is assumed for a disc with a constant rotation velocity and $Q_{crit} = 1$ for a quasi-stable disc as identified in many studies (e.g. \citeauthor{ForsterSchreiber2006}  \citeyear{ForsterSchreiber2006}, \citeauthor{Genzel2011} \citeyear{Genzel2011}, \citeauthor{Burkert2010} \citeyear{Burkert2010}).

%--------------------------------------------
%Figure Dispersion VS redshift
   \begin{figure}
   \centering
   \includegraphics[scale=0.385]{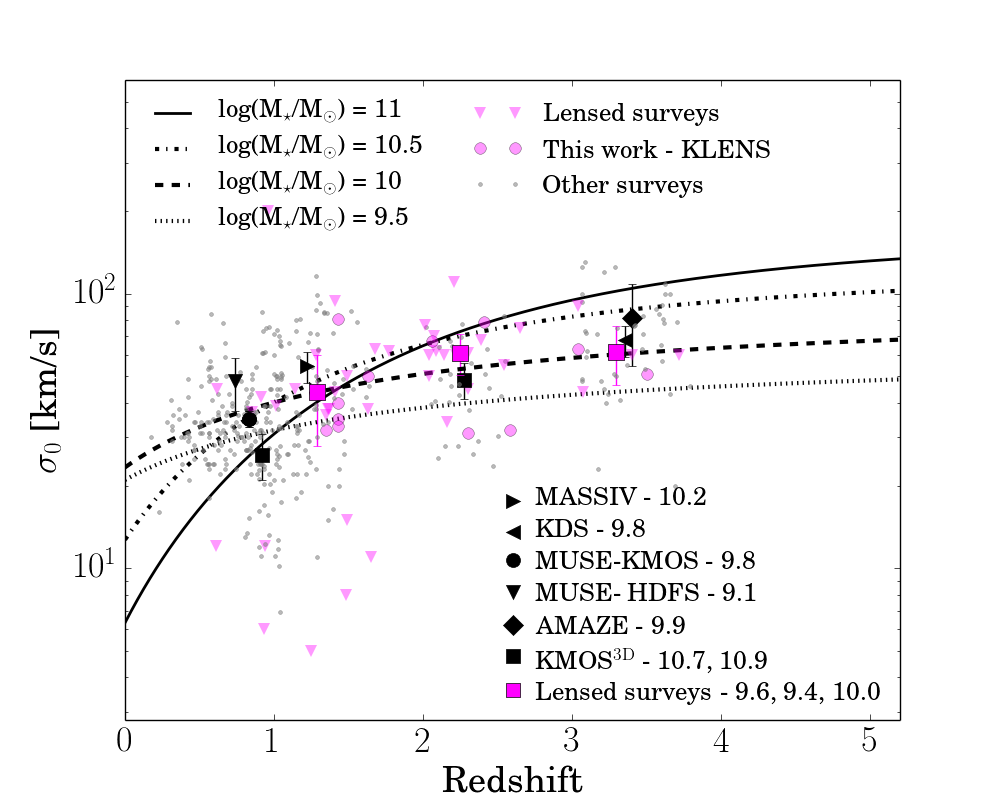}
   \caption{Intrinsic velocity dispersion as a function of redshift. The curves are from Eq. (\ref{sigma_final_2}) for $\rm log(M_{\star}/M_\odot)= $ 9.5, 10.0, 10.5, and 11.0. The magenta squares represent the mean velocity dispersions of the lensed surveys, which are the combination of our KLENS work,  \citet{Livermore2015}, \citet{Leethochawalit2016}, and KLASS \citep{Mason2017}. The lensed surveys are divided in three redshift bins : $z<2$, $2<z<3$, and $z>3$. The black symbols present the mean velocity dispersions of the surveys listed at the lower right with their corresponding mean stellar masses. 
   }
\label{sigma_vs_z}%
   \end{figure}
   %--------------------------------------------

\cite{Reyes2011} have constrained the Tully-Fisher relation for the local galaxies :

\begin{ceqn}
\begin{align}
\label{TF_Reyes2011}
\mathrm{log} \, \upsilon_{rot}(\mathrm{M}_{\star}) = 2.127 + 0.278 \, ( \mathrm{log}  \, \mathrm{M}_{\star} - 10.10) \, \,  \mathrm{[km \, s^{-1}]}.
\end{align}
\end{ceqn}

Many studies try to understand and constrain the evolution of this relation at higher redshifts, but recent studies show disparate results (e.g \citeauthor{Straatman2017} \citeyear{Straatman2017}; \citeauthor{Ubler2017} \citeyear{Ubler2017}; \citeauthor{Harrison2017} \citeyear{Harrison2017}). To obtain a relation of the velocity dispersion strictly as a function of the redshift and stellar mass, we can rewrite Eq. (\ref{eq_sigma}) replacing the rotation velocity $\upsilon_{rot}$ by the Tully-Fisher relation (Eq. \ref{TF_Reyes2011}) assuming it does not evolve with redshift :

\begin{ceqn}
\begin{align}
\label{sigma_final_2}
\sigma_0(z, \mathrm{M}_{\star}) \simeq 0.147464 \, \mathrm{M}_{\star}^{0.278} \, f_{gas}(z, \mathrm{M}_{\star}) \, \,  \mathrm{[km \, s^{-1}]},
\end{align}
\end{ceqn}
where $f_{gas}$ is given by Eqs (\ref{eq1}-\ref{eq2}). From Eq. (\ref{sigma_final_2}), we can study the relation between the velocity dispersion and redshift, the $\sigma_0-z$ relation, at a fixed stellar mass, and the relation between velocity dispersion and stellar mass, the $\rm \sigma_0-M_\star$ relation, at fixed redshift.

Figure \ref{sigma_vs_z} shows the predicted velocity dispersion as a function of redshift. The curves are defined from Eq. (\ref{sigma_final_2}) where we fix $\rm log(M_\star/M_\odot) $ at 9.5, 10.0, 10.5, and 11.0. A relatively steep evolution is found for the higher mass model, whilst a shallower evolution is found as mass decreases. We include in this figure measures from the different surveys mentioned in Sect. \ref{comparisonsamples}. We divide the lensed surveys in three different redshift bins : $z<2$, $2<z<3$, and $z>3$, since they include galaxies mostly at $1<z<3.5$. The respective $\rm log(M_{\star}/M_\odot)$ mean values of the latter galaxies are 9.6, 9.4, and 10 for each redshift bin. The lensed surveys now allow us to test the observed trend of the dispersion as a function of redshift for the lower mass galaxies and add a significant number of galaxies, especially at $z>1$.

We can see the values from the lensed surveys, as well as all the other surveys, are overall in good agreement with the theoretical curves (Eqs \ref{eq1}-\ref{sigma_final_2}) at their corresponding mass, taking into account the spread of the measured values in each redshift bin. We have to keep in mind here that the values from the different samples have not been corrected in the same way for the beam smearing. The lensed surveys have several galaxies for which we measure an upper limit on $\sigma_0$, which means that the obtained values could be higher than the intrinsic velocity dispersions. If we apply an extra correction for the beam smearing, we obtain that our measurements are still in agreement with the model since the expected corrections are small for low mass galaxies and the errors due to the spread of the values in each redshift bin are larger than the correction \citep[see Sect. \ref{section52}; e.g.][]{Burkert2016, Johnson2017}.

As a result, globally the observations follow well the evolution model and interpretation of the intrinsic velocity dispersion as a function of redshift proposed by \citet{Wisnioski2015}. We can now push further the analysis and try to distinguish the evolution of massive galaxies compared to low mass galaxies.

%-------------------------------------------------------
%Figure Dispersion VS mass
   \begin{figure}
   \centering
   \includegraphics[scale=0.385]{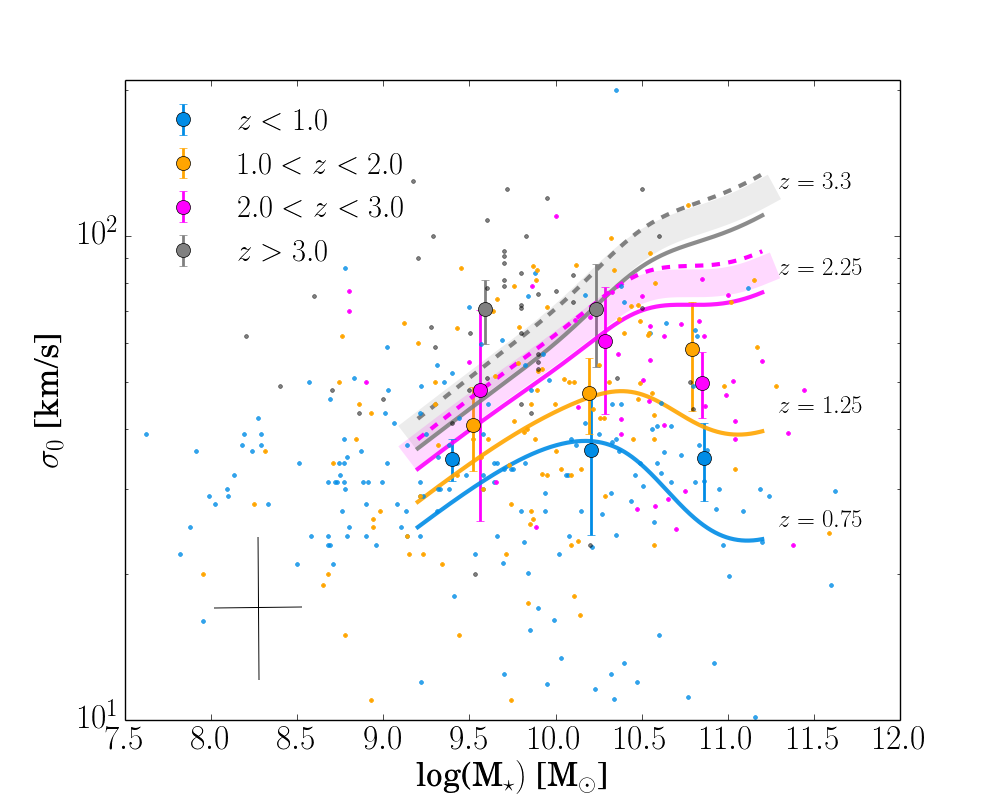}
   \caption{Intrinsic velocity dispersion as a function of stellar mass. The solid and dashed lines are the $\rm \sigma_0-M_\star$ relations from Eq. (\ref{sigma_final_2}) and Eq. (\ref{sigma_final_3}) derived with the Tully-Fisher relation of \citet{Reyes2011} and \citet{Straatman2017}, respectively. The redshifts correspond to 0.75, 1.25, 2.25 and 3.3, the mean values of each redshift bin. The circles represent the mean velocity dispersions of each stellar mass bin: $\rm 8.9 < log(M_{\star}/M_\odot)<9.9$, $\rm 9.9 < log(M_{\star}/M_\odot)<10.5$, and $\rm  log(M_{\star}/M_\odot)>10.5$. The data are from the same surveys as in Fig. \ref{sigma_vs_z} when the mass is available. The typical error on individual datapoints is presented at the bottom left.}
              \label{sigma_vs_mass}%
   \end{figure}
   
%-------------------------------------------------------
   % Figure histograms dispersion for low and high mass
         \begin{figure}
   \centering
  \subfloat{\includegraphics[scale=0.365]{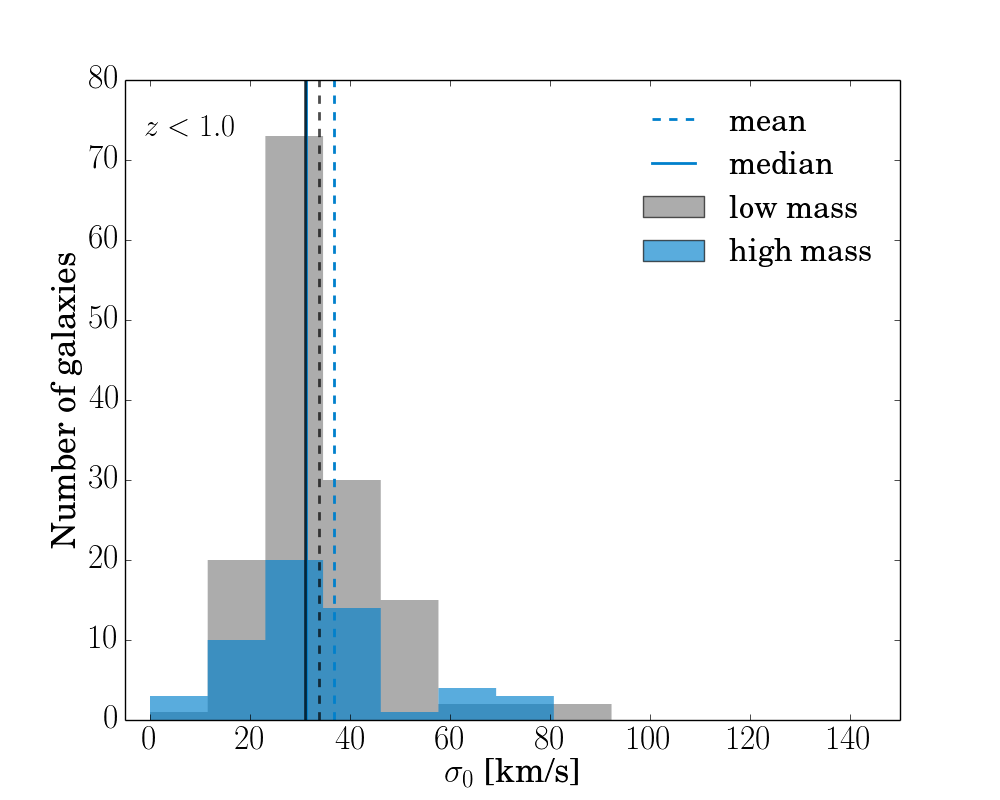}}\\
   \subfloat{\includegraphics[scale=0.365]{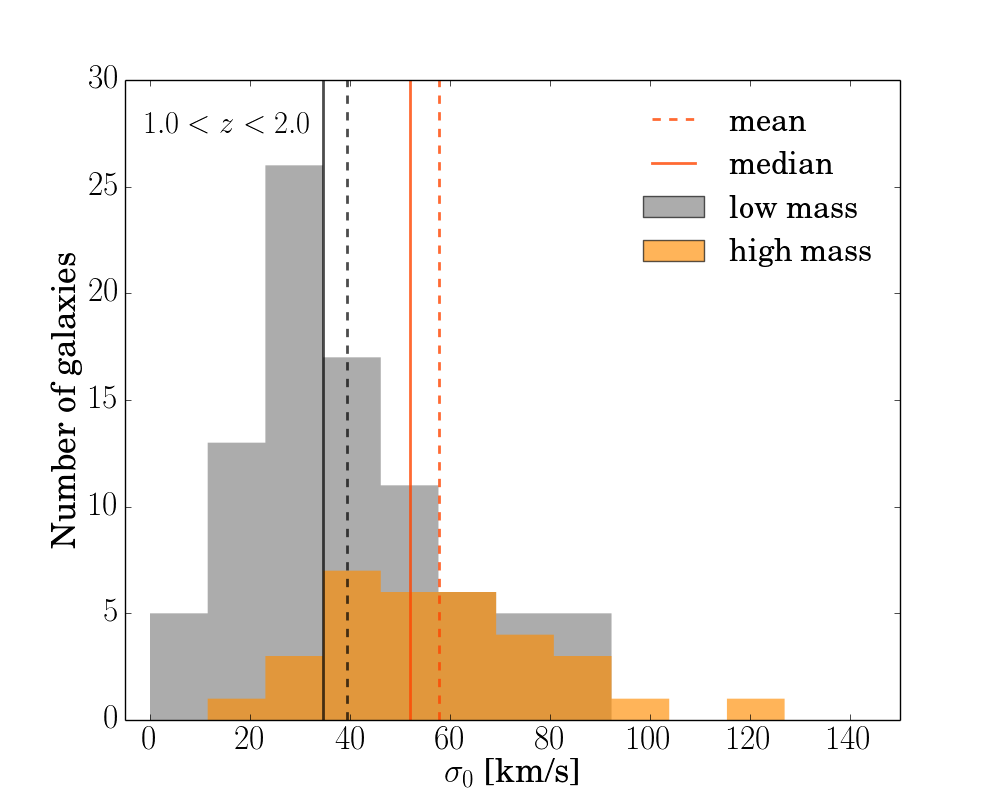}}\\
  \subfloat{\includegraphics[scale=0.365]{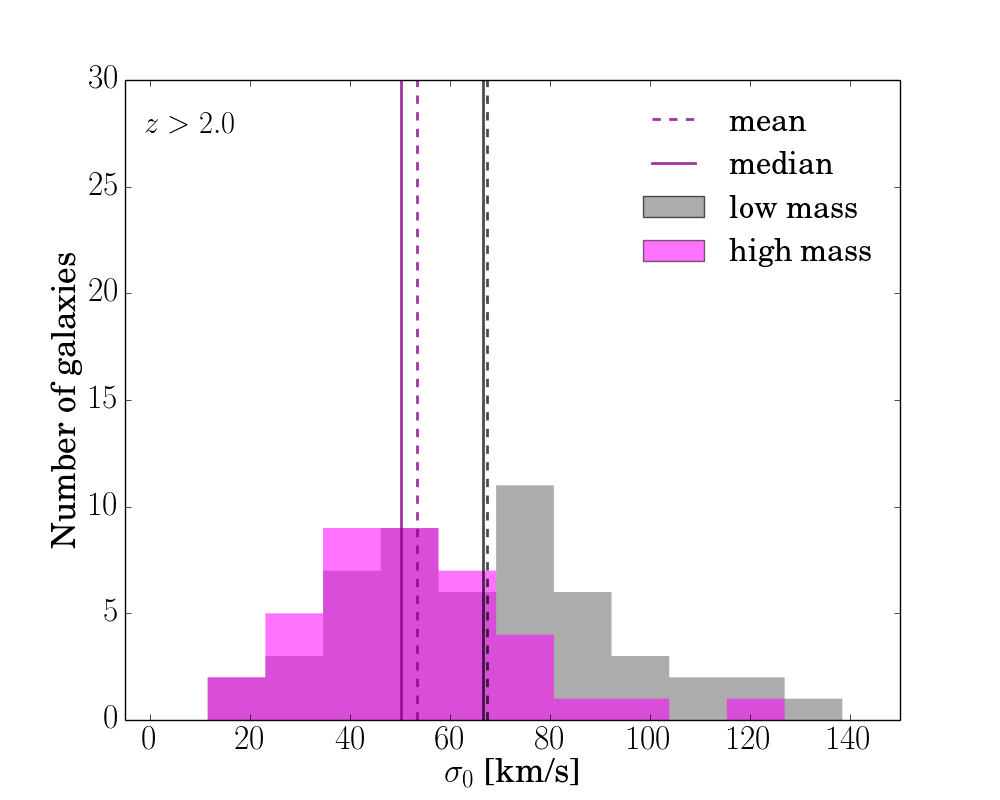}}

      \caption{Histograms of the intrinsic velocity dispersion for low and high masses at $z<1.0$ (top panel), $1.0<z<2.0$ (middle panel) and $z>2$ (bottom panel). The low mass galaxies are defined as $\rm log( M_{\star}/M_\odot)<10.2$ and high mass as $\rm log( M_{\star}/M_\odot)>~10.2$. The compilation is the same one as in Fig.~\ref{sigma_vs_mass}.}
         \label{hist_disp}
   \end{figure}
%-------------------------------------------------------

The dispersion as a function of stellar mass is presented in Fig. \ref{sigma_vs_mass}. The solid lines assumes the Tully-Fisher relation of \citet{Reyes2011} described by Eq. (\ref{sigma_final_2}), where we fix the redshift at the mean values of each redshift bin, i.e. 0.75, 1.25, 2.25, and 3.3. The redshift bins are defined in the same way as in Fig.~\ref{sigma_vs_z}, except that we add $z<1$ and $1<z<2$ now that we combine all the surveys together and have many galaxies in these bins. The dashed lines at $z=2.25$ and $z=3.3$ use the Tully-Fisher relation obtained by \citet{Straatman2017} at $2.0<z<2.5$, which is probably more appropriate for these redshifts :

\begin{ceqn}
\begin{align}
\label{sigma_final_3_1}
 \rm log \, \upsilon_{rot}(\mathrm{M}_{\star}) = 2.17 + 0.29 \, ( \mathrm{log} \, M_{\star} - 10.0) \, \,  \mathrm{[km \, s^{-1}]}.
\end{align}
\end{ceqn}

We thus obtain the $\rm \sigma_0-M_\star$ relation rewriting Eq. (\ref{eq_sigma}), this time using Eq. (\ref{sigma_final_3_1}) :

\begin{ceqn}
\begin{align}
\label{sigma_final_3}
\sigma_0(z, \mathrm{M}_{\star}) \simeq 0.131669 \, \mathrm{M}_{\star}^{0.29} \, f_{gas}(z, \mathrm{M}_{\star})  \, \,  \mathrm{[km \, s^{-1}]}.
\end{align}
\end{ceqn}

At $z\gtrsim1$, the plotted curves predict lower intrinsic velocity dispersions for lower masses, and an increase for higher masses, stronger at higher redshift due to the rapid decrease of the gas fraction with stellar mass. At $z\lesssim1$, the curves predict lower intrinsic velocity dispersions for the highest masses. 

We consider three stellar mass bins for the plotted samples, $\rm 8.9 < log(M_{\star}/M_\odot)<9.9$, $\rm 9.9 < log(M_{\star}/M_\odot)<10.5$, and $\rm  log(M_{\star}/M_\odot)>10.5$. Globally, we find a clear trend showing higher dispersion at high redshift for low and high mass galaxies as expected. The mean dispersions of each stellar mass bin are in good agreement with the curves within their 1.5$\sigma$ dispersion, except for galaxies with a mass of $\rm 8.9 < log(M_{\star}/M_\odot)<9.9$ and  $\rm  log(M_{\star}/M_\odot)>10.5$ which show  a discrepancy at higher redshift with higher and lower velocity dispersions than expected from the above relations, respectively (discussion below).

To better appreciate what is going on, Fig. \ref{hist_disp} presents histograms of the velocity dispersion for the same data as in  Fig. \ref{sigma_vs_mass}, but we divide here the galaxies in two sub-samples with high masses, $\rm log( M_{\star}/M_\odot) > 10.2$, and low masses,  $\rm log( M_{\star}/M_\odot)~<~10.2$. The redshift bins are defined in the same way as in Fig. \ref{sigma_vs_mass}, apart from the higher redshifts for which we combine all the galaxies at $z>2$ in only one bin, since most of the low mass galaxies are at $z>3$ while most of the high mass are between $2<z<3$ (therefore we do not have enough low and high mass galaxies at high redshift to divide our sample in two different redshift bins).

At $ z<1$, we obtain a similar velocity dispersion median of $\rm 31 \pm 3 \, km \, s^{-1} $ and $\rm 31 \pm 8 \, km \, s^{-1} $ for the low and high mass galaxies, respectively ($z_{mean} \sim0.75$; Fig. \ref{hist_disp}, top panel). This is expected at this redshift since the relations derived from Eq. (\ref{sigma_final_2}) have comparable values regardless of the mass. At $1<z<2$, we obtain a median of the velocity dispersion of $\rm 35 \pm 5 \, km \, s^{-1} $ and  $\rm 52 \pm 9 \, km \, s^{-1} $ for the low and high mass galaxies, respectively ($z_{mean} \sim1.25$; Fig. \ref{hist_disp}, middle panel). This result is in agreement with the  $\rm \sigma_0 - M_\star$ relation and can be explained by the redshift evolution  obtained at this redshift range for high mass galaxies, yielding a more rapid increase in dispersion for high mass than low mass galaxies. 

At $z>2$, however, we obtain a median of $\rm 66 \pm 9 \, km \, s^{-1} $ and $\rm 50 \pm 8 \, km \, s^{-1} $ for the low and high mass galaxies, respectively, when from the $\rm \sigma_0 - M_\star$ relation we expect to obtain values of $\rm \sim~51 \, km \, s^{-1}$ and $\rm \sim 81 \, km \, s^{-1} $. Figure \ref{hist_disp} (bottom panel), in addition to Fig. \ref{sigma_vs_mass}, supports that low and high mass galaxies at $z>2$ do not seem to follow the established $\rm \sigma_0 - M_\star$ relations. The velocity dispersions found at high redshift seem to be higher than expected for low mass galaxies, and lower for high mass galaxies.

The beam smearing effect could play a role in the results obtained here. Indeed, if the measured velocity dispersions are higher than the intrinsic velocity dispersion, it could explain why we obtain higher values compared to the curves in many cases (see the datapoints at redshifts of $0.75$ and $1.25$ in Fig. \ref{sigma_vs_mass}). However, it cannot explain the lower value of $\sigma_0$ we obtain at $z>2$ for the high mass galaxies, since a beam smearing correction would have the effet to lower the value even more and increase the observed discrepancy. 

The velocity dispersion discrepancy at high redshift could be caused by an observational bias. It is possible that we only observe the low mass galaxies which have the higher dispersions because they are the easier to measure. We also might not be probing a representative sample of galaxies, in terms of their physical properties, at this epoch.
In fact, as mentioned before, in the histogram showing the redshift bin $z~>~2$ (Fig. \ref{hist_disp}, bottom panel), most of the low mass galaxies are at $ z>3$ ($z_{mean} \sim 3.15$) while the majority of the high mass galaxies are between $ 2<z<3$ ($z_{mean} \sim 2.4$). 

As proposed by some studies (e.g. \citeauthor{Ubler2017} \citeyear{Ubler2017}), it is also possible that the Tully-Fisher relation evolves more strongly at higher redshift, which would cause an offset toward higher dispersion, as we see in Fig.\ref{sigma_vs_mass} with the Tully-Fisher from \citet{Straatman2017}. However, we observe this offset only for the low mass galaxies, while the galaxies with masses of $\rm 9.9<log(M_\star/M_\odot)<10.5 $ seem to be in better agreement with the $\rm \sigma_0-M_\star$ relation from Eq. (\ref{sigma_final_2}). Moreover, a stronger evolution of the Tully-Fisher relation does not solve the discrepancy of the observations at the high-mass end, on the contrary. An evolution of the slope of the Tully-Fisher relation would also directly affect the slope of the $\rm \sigma_0-M_\star$ relation.

In the assumption of the established relations (Eqs \ref{eq1}-\ref{sigma_final_2}), the sSFR has been determined for a valid range of $0.5<z<2.5$. Using the $\rm SFR-M_\star$ relation of \cite{Tomczak2016}, parametrized up to $z=4$, we obtain slightly different curves than those shown in Fig. \ref{sigma_vs_mass}. The velocity dispersions obtained for massive galaxies ($\rm log(M_\star/M_\odot) >10.5$) are quasi-constant at $z > 2$, with lower predicted values. However, the predicted velocity dispersions are still not low enough to be consistent with the observed velocity dispersions at this redshift.

The disagreement between data and the model at $z>2$ could be an indication that the model with $Q_{crit}=1$ fails for galaxies in these redshift and/or mass bins.
A value of $Q_{crit} >1$ would be necessary to be in agreement with the high velocity dispersion observed for low mass galaxies at $ z>2$, while a $Q_{crit} <1$ would be necessary to obtain smaller velocity dispersion for high mass galaxies at the same redshift. A value of $ Q_{crit} <1$ usually is an indication of an unstable disc, while $ Q_{crit}>1$ indicates a stable disc \citep{Genzel2011}.

%%%%%%%%%%%%%%%%%%%%%%%%%%%%%%%%%%%%%%%%
%           Conclusion
%%%%%%%%%%%%%%%%%%%%%%%%%%%%%%%%%%%%%%%%

\section{Conclusions}
\label{Conclusions}

We have presented KLENS, a survey exploiting gravitational lensing to study SFGs with low masses and/or SFR using the KMOS near-IR multi-object spectrograph at the VLT. We here report the first results using H and K band observations of two lensing clusters : AS1063 and MACS1206-08. We detect emission lines in 24 galaxies at $1.4<z<3.5$ with a median mass of $\rm log(M_\star/M_\odot)=9.6$ and median SFR of 7.5 $\rm M_\odot \, yr^{-1}$. We show a morpho-kinematic analysis of our sample, and compare our kinematic classification with other surveys from the literature. We combine several surveys together to investigate the evolution of the intrinsic velocity dispersion as a function of the redshift and stellar mass as well as the dependence between the fraction of rotation dominated galaxies and integrated properties. The main results from this paper are summarized as follows.

   \begin{enumerate}

    \item There is a clear velocity gradient in 12 objects in our sample, while the 12 other are more compact, clumpy or show irregular kinematics. For most of the objects showing a velocity gradient, we do not observe the flattening of the rotation curves or a peak in the velocity dispersion map at the kinematic centre. Eight of them have $\rm \Delta PA < 30^\circ$ and two seem to show on-going interaction.

    \item From our classification, we obtain a fraction of 25\% of galaxies dominated by rotation. Using the criterion of \citet{ForsterSchreiber2009}, we obtain 40\%. 
      
      \item When combining our data with other surveys, we find that the fraction of rotation dominated galaxies appears to increase with stellar mass, while it decreases with a positive offset from the star-formation main sequence. 
      
      \item We derive  $z-\sigma_0$ and $\rm M_\star-\sigma_0 $ relations using the Tully-Fisher relation and the equations previously obtained for the velocity dispersion by \citet{Wisnioski2015}. We assume that galaxies are in quasi-equilibrium (Toomre Q parameter equal to 1).
      
      \item From the $z-\sigma_0$ relation, we find a stronger redshift evolution of the velocity dispersion for massive galaxies.
      The observations are overall in good agreement with the $z-\sigma_0$ relation corresponding to their mass.
      
   \item  From the $\rm M_\star-\sigma_0 $ relation, we obtain lower velocity dispersions for lower stellar masses, and an increase for higher masses, stronger at higher redshift. Observations are consistent with these relations, except at $z>2$, where we observe higher velocity dispersion for low mass galaxies and lower velocity dispersions for high mass galaxies. This disagreement between theoretical curves and observations could suggest that the Tully-Fisher relation that we use is inadequate for this mass and/or redshift range. It could also indicate that the classical Toomre stability criterion is not satisfied at high redshift (with the Toomre Q parameter different from 1), or that the adopted parametrisation of the sSFR and molecular properties fail at high redshift.

\end{enumerate}

\begin{acknowledgements}
\\

This work has made use of the Rainbow Cosmological Surveys Database, which is operated by the Universidad Complutense de Madrid (UCM), partnered with the University of California Observatories at Santa Cruz (UCO/Lick,UCSC). 
We are very grateful to Tucker Jones and Nicha Leethochawalit for sharing their 2D modeling codes. We also want to thank  Mark Swinbank for advices on the data reduction and Nicolas Bouch\'{e} for discussions on GalPaK$\rm ^{3D}$.
This work was supported by the Swiss National Science Foundation.
MG is grateful to the Fonds de recherche du Qu\'{e}bec - Nature et Technologies (FRQNT) for  the financial support. PGP-G acknowledges support from Spanish Government MINECO grants AYA2015-63650-P and AYA2015-70815-ERC.
\end{acknowledgements}

\bibliographystyle{aa} % 
\bibliography{references} % 

\appendix

\section{Kinematic maps}
\label{appA}

Figures \ref{image1} and \ref{image3}  show the HST images, H$\alpha$ or [OIII] flux, the velocity and velocity dispersion maps of each galaxy of our sample.
Figure \ref{image5} presents the observed velocity maps, the velocity maps from the models, the rotation curves and the velocity dispersion profiles of each galaxy showing a velocity gradient.

%-------------------------------------------------------
%Figure 
   \begin{figure*}
   \centering
   \subfloat{\includegraphics[scale=0.6]{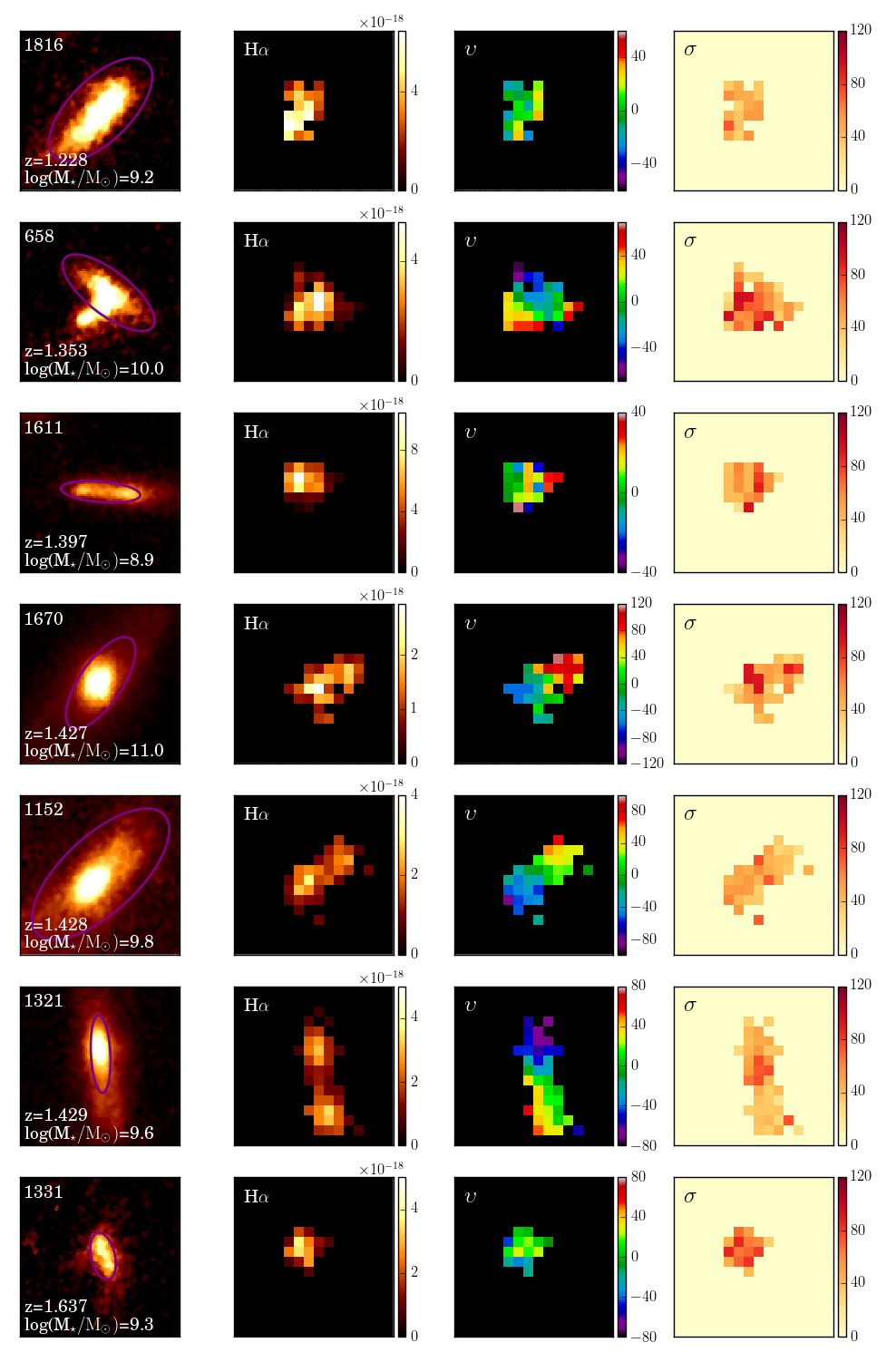}}
   \caption{The HST/WFC3 F160W near-infrared images, the H$\alpha$ or [OIII] flux in $erg \, s^{-1} \, cm^{-1}$, the velocity and velocity dispersion maps in $ km \, s^{-1}$ of galaxies from the cluster AS1063. The ellipses in the HST/WFC3 images represent the effective radii.  }
              \label{image1}%
   \end{figure*}

    \begin{figure*} 
   \centering
    \setcounter{figure}{0}
    \subfloat{\includegraphics[scale=0.6]{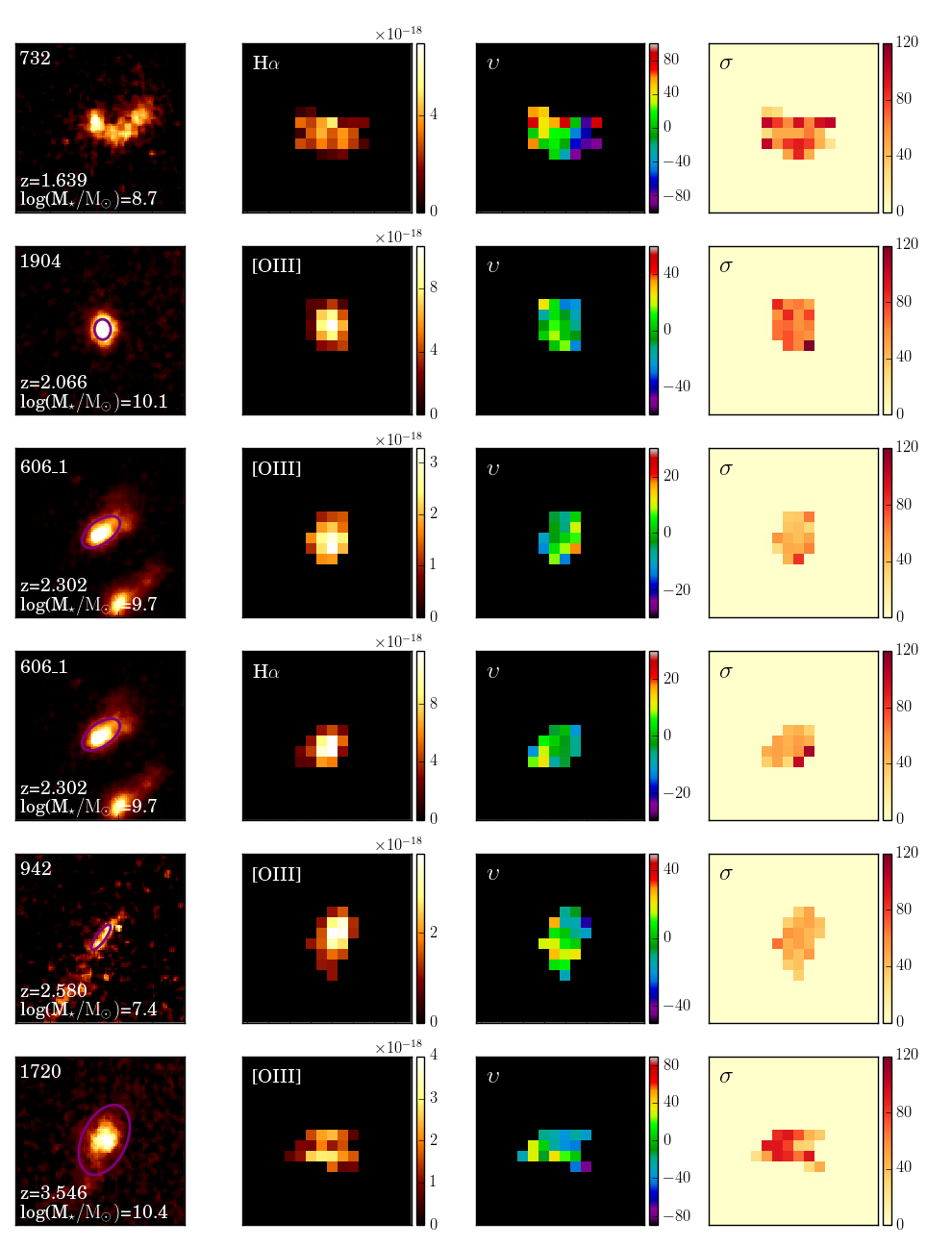}}
   \caption{Continued.}
              \label{image2}%
   \end{figure*}

   \begin{figure*}
   \centering
 \includegraphics[scale=0.6]{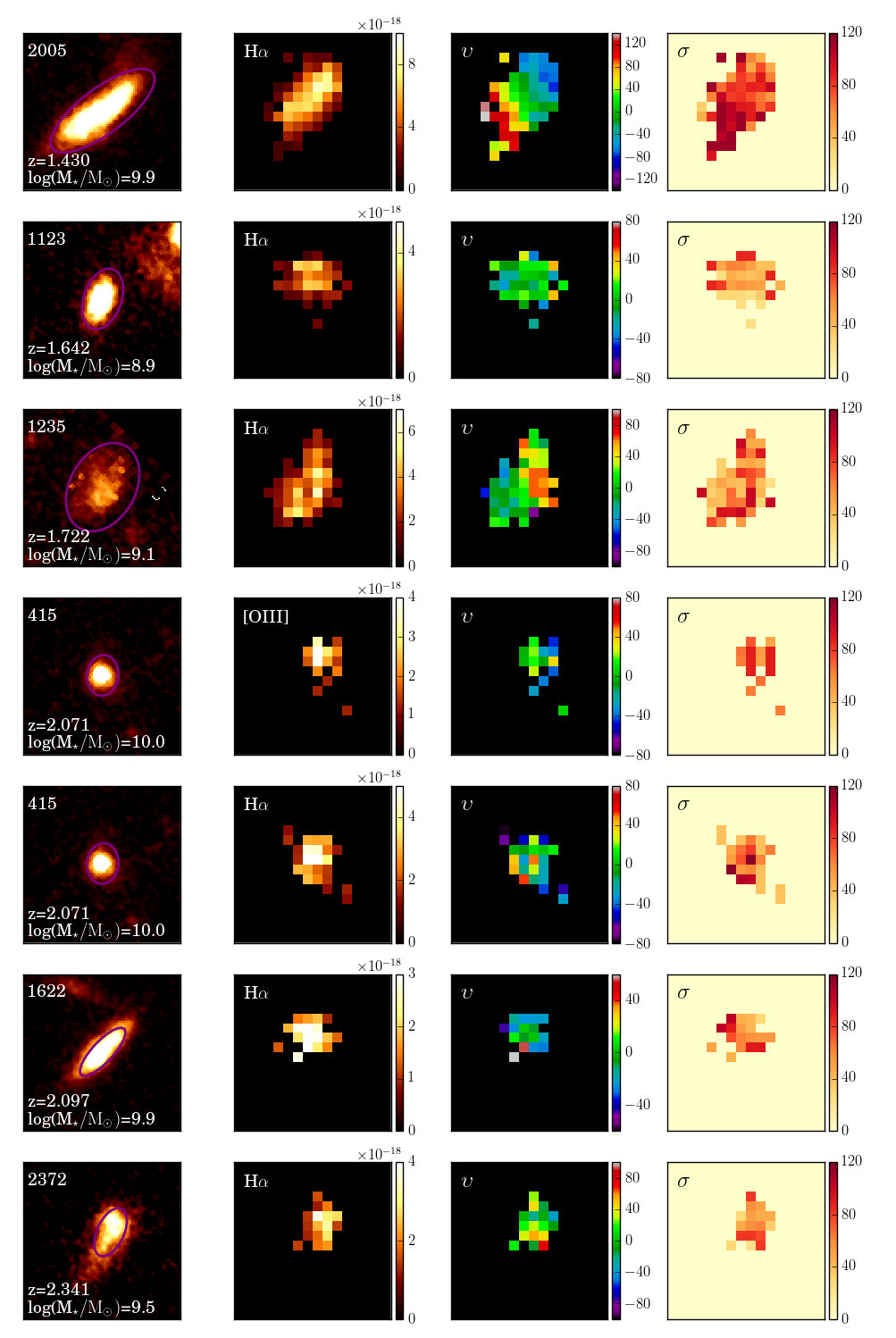}
   \caption{Same as Fig. \ref{image1} for galaxies from MACS1206-08. }
              \label{image3}%
   \end{figure*}

  \begin{figure*}
   \centering
    
    \includegraphics[scale=0.6]{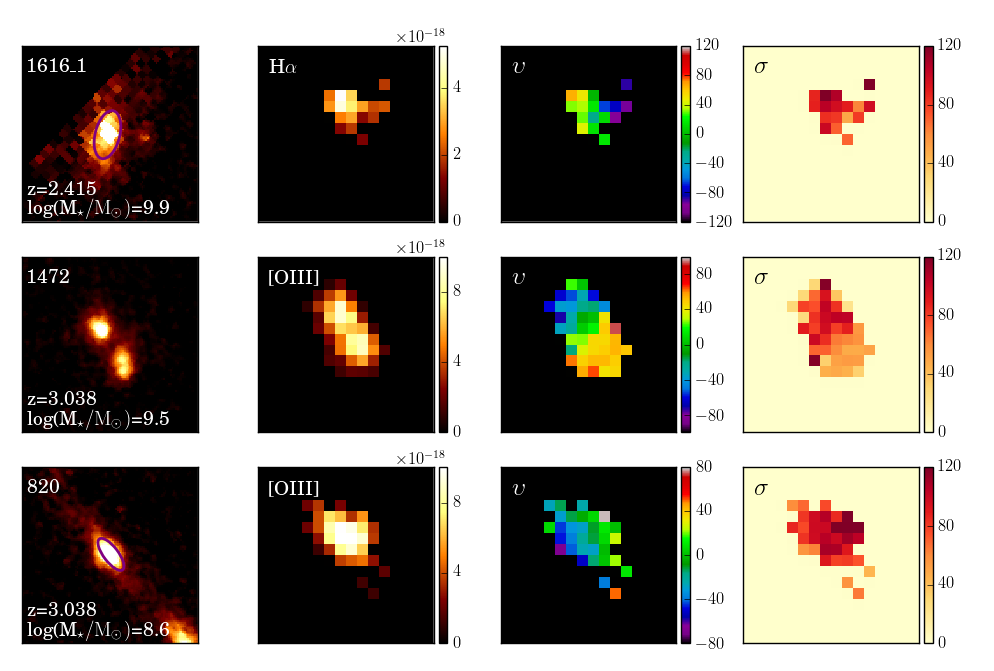}
\setcounter{figure}{1}
   \caption{Continued. }
              \label{image4}%
   \end{figure*}

   \begin{figure*}
   \centering
  \includegraphics[scale=0.59]{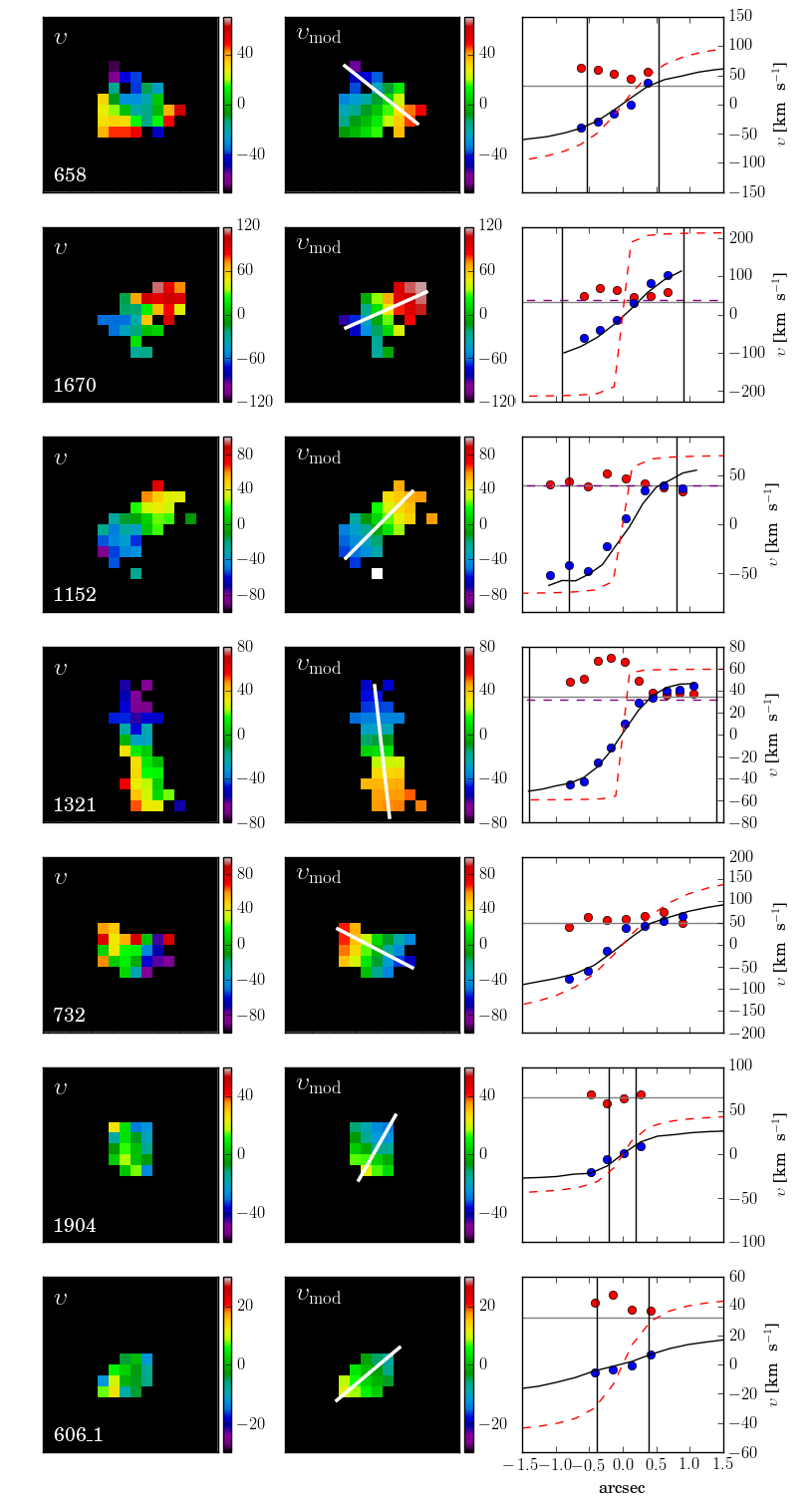}
   \caption{Observed velocity maps, velocity maps from the kinematic models and rotation curves extracted on the major axis as indicated by the white solid lines on the maps. The blue circles and black solid line represent the curve extracted from the observed velocity maps and velocity maps from the model, respectively. The red dashed lines show the intrinsic rotation curves from the models. The vertical black solid lines indicate the value of the effective radius $R_e$. The red circles and the horizontal black solid lines represent the velocity dispersion profile and the measured intrinsic velocity dispersion. The horizontal purple dashed lines indicate the intrinsic velocity dispersion obtained with GalPaK$^{\rm 3D}$ for 5 galaxies.  }
              \label{image4}%
   \end{figure*}

  \begin{figure*}
   \centering
   
    \includegraphics[scale=0.595]{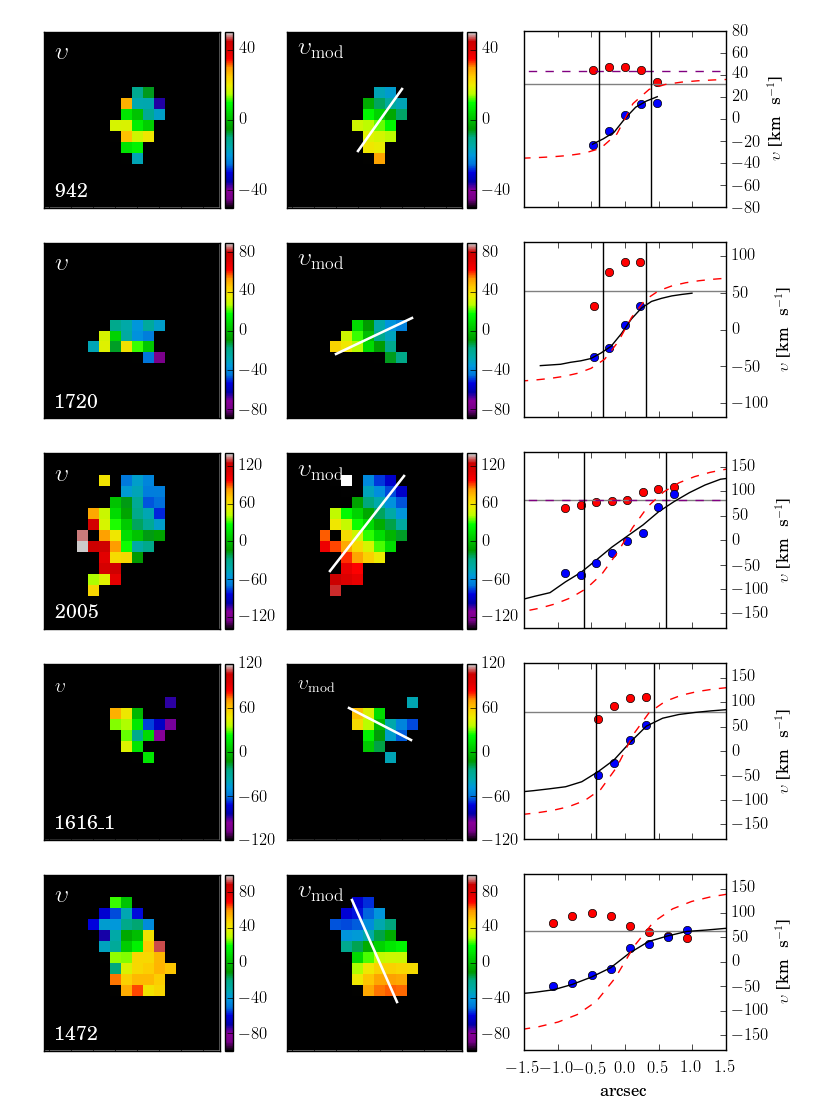}
\setcounter{figure}{2}
   \caption{Continued.}
              \label{image5}%
   \end{figure*}

\end{document}